\newcommand\T{\rule{0pt}{2.6ex}}
\newcommand\B{\rule[-1.2ex]{0pt}{0pt}}
\documentclass[UKenglish, cernpreprint, texmf]{atlasdoc}
 
\usepackage{atlaspackage}
\usepackage{multirow}
\usepackage{atlasbiblatex}
 
\usepackage{atlasphysics}
 
\graphicspath{{logos/}{figures/}}
 
\usepackage{float}
\PreprintIdNumber{CERN-EP-2020-077}
\AtlasJournal{JHEP}
\LEcontact{Bob Kowalewski bob.kowalewski@cern.ch}
 
\AtlasTitle{Measurement of hadronic event shapes in high-$p_{\textrm{T}}$ multijet final states at $\sqrt{s} = 13$ TeV with the ATLAS detector}
 
\AtlasVersion{3.0}
\AtlasJournal{JHEP}
\AtlasJournalRef{JHEP 01 (2021) 188}
\AtlasDOI{10.1007/JHEP01(2021)188}
\AtlasAbstract{
A measurement of event-shape variables in proton--proton collisions at large momentum transfer is presented using data collected at $\sqrt{s} = 13$ TeV with the ATLAS detector at the Large Hadron Collider. Six event-shape variables calculated using hadronic jets are studied in inclusive multijet events using data corresponding to an integrated luminosity of 139~\ifb. Measurements are performed in bins of jet multiplicity and in different ranges of the scalar sum of the transverse momenta of the two leading jets, reaching scales beyond 2 TeV. These measurements are compared with predictions from Monte Carlo event generators containing leading-order or next-to-leading order matrix elements matched to parton showers simulated to leading-logarithm accuracy. At low jet multiplicities, shape discrepancies between the measurements and the Monte Carlo predictions are observed. At high jet multiplicities, the shapes are better described but discrepancies in the normalisation are observed.
}
 
\author{The ATLAS Collaboration}
 
\AtlasRefCode{STDM-2019-02}
 
\AtlasDate{\today}

\hypersetup{pdftitle={ATLAS document},pdfauthor={The ATLAS Collaboration}}
 
\begin{document}
 
\maketitle
 
\tableofcontents
 
\newpage
\section{Introduction}
\label{sec:intro}
Event shapes \cite{evshapes1, evshapes2} are a class of observables that describe the dynamics of energy flow in multijet final states. Normally, event-shape observables are defined such that they vanish for 2$\rightarrow$ 2 processes and increase to a maximum for final states with uniformly distributed energy. These observables are sensitive to different aspects of the theoretical description of these strong-interaction processes. They are usually defined to be infrared and collinear safe, which enables their calculation in perturbative Quantum Chromodynamics (QCD). Hard, wide-angle radiation is studied by investigating the tails of the event-shape distributions. These configurations are sensitive to higher-order corrections to the dijet cross section, which are available up to next-to-next-to-leading order (NNLO) \cite{dijetNNLO}.
Other regions of the event-shape distributions provide information about anisotropic, back-to-back configurations, which are sensitive to the details of the resummation of soft logarithms in the theoretical predictions. \\
\newline
Event-shape observables have been measured in $e^{+} e^{-}$ collisions at LEP \cite{esAleph, esOPAL, esDELPHI, esL3}, $ep$ collisions at HERA \cite{esZEUS}, and $p\bar{p}$ collisions at the Tevatron \cite{esCDF}. More recently, they have been measured in $pp$ collisions at $\sqrt{s} = 7$ TeV by the ALICE, CMS and ATLAS Collaborations \cite{esAlice, esCMS7, CMS-EWK-11-021, esAtlas}. ALICE also published measurements at $\sqrt{s} = 0.9$ and 2.76 TeV \cite{esAlice}, and the CMS Collaboration published a measurement at $\sqrt{s} = 13$ TeV \cite{esCMS13}.\\
\newline
In this study, several different event-shape variables are investigated, probing the properties of the multijet energy flow at the large, $O$(TeV), energy scales provided by the Large Hadron Collider (LHC) at $\sqrt{s}=13$ TeV. Measurements are compared with fixed-order matrix elements matched to parton shower Monte Carlo (MC) predictions for a selection of observables that cover various aspects of the physics of multijet processes. These observables are defined in detail in Section~\ref{sec:strategy}. The measurements are performed for different energy regimes, given by the scalar sum of transverse momenta of the two leading jets, $H_{\mathrm{T}2} = p_{\mathrm{T}1}+p_{\mathrm{T}2}$, and the jet multiplicity, $n^{\textrm{jet}}$. The phase-space region explored in this analysis is defined by $H_{\mathrm{T}2} > 1~\TeV$, with jet $\pt > 100~\GeV$ to reduce experimental uncertainties and non-perturbative effects. This paper extends the currently available measurements by studying the dependence of event shapes on $n^{\textrm{jet}}$, which is not usually found in the literature. This approach provides inputs with which to compare future higher-order QCD predictions, since the power of $\alpha_\text{s}$ in the perturbative expansion of the cross section increases with each additional jet emission in the final state. In addition, it allows the definition of phase-space regions sensitive to processes beyond-the-Standard-Model, which are often characterised by isotropically distributed multijet final states~\cite{LED,UED,SUSY}. Measurements of the differential cross sections as a function of $n^{\textrm{jet}}$ for different energy scales are also reported.\\
\newline
The paper is organised as follows. The ATLAS detector is described in Section~\ref{sec:detector}. The measurement strategy and the definitions of the observables are discussed in Section~\ref{sec:strategy}, followed by the details of the data sample and MC simulations in Section~\ref{sec:MC_simulation}. Section~\ref{sec:data_selection} is dedicated to the object and event-selection criteria. The correction to particle level is described in Section~\ref{sec:unfolding}, followed by a discussion of systematic uncertainties in Section~\ref{sec:syst_uncert}. The particle-level results are compared with MC predictions in Section~\ref{sec:result} and the summary and conclusions are provided in Section~\ref{sec:conclusions}.
\section{ATLAS detector}
\label{sec:detector}
\newcommand{\AtlasCoordFootnote}{ATLAS uses a right-handed coordinate system with its origin at the nominal interaction point (IP) in the centre of the detector and the \(z\)-axis along the beam pipe. The \(x\)-axis points from the IP to the centre of the LHC ring, and the \(y\)-axis points upwards. Cylindrical coordinates \((r,\phi)\) are used in the transverse plane, \(\phi\) being the azimuthal angle around the \(z\)-axis. The pseudorapidity is defined in terms of the polar angle \(\theta\) as \(\eta = -\ln \tan(\theta/2)\).}
The ATLAS detector~\cite{PERF-2007-01} at the LHC covers nearly the entire solid angle around the collision point.\footnote{\AtlasCoordFootnote}
It consists of an inner charged-particle tracking detector surrounded by a thin superconducting solenoid, electromagnetic and hadronic calorimeters, and a muon spectrometer incorporating three large superconducting toroidal magnets.\\
\newline
The inner-detector system is immersed in a \SI{2}{\tesla} axial magnetic field and provides charged-particle tracking in the range \(|\eta| < 2.5\).
Closest to the interaction point, the high-granularity silicon pixel detector covers the vertex region and typically provides four measurements per track, with the first hit normally recorded in the insertable B-layer installed before Run~2~\cite{ATLAS-TDR-2010-19,PIX-2018-001}. It is followed by the silicon microstrip tracker, which usually provides eight measurements per track. These silicon detectors are complemented by the transition radiation tracker (TRT), which enables radially extended track reconstruction up to \(|\eta| = 2.0\). The TRT also provides electron identification information based on the fraction of hits (typically 30 in total) above a high energy-deposit threshold that corresponds to transition radiation.\\
\newline
The calorimeter system covers the pseudorapidity range \(|\eta| < 4.9\). Within the region \(|\eta|< 3.2\), electromagnetic calorimetry is provided by barrel and endcap high-granularity lead/liquid-argon (LAr) calorimeters, with an additional thin LAr presampler that covers \(|\eta| < 1.8\), to correct for energy loss in material upstream of the calorimeters. Hadronic calorimetry is provided by the steel/scintillator-tile calorimeter, segmented into three barrel structures in the region \(|\eta| < 1.7\), and two copper/LAr calorimeters in the endcap regions (\(1.5 <|\eta|< 3.2\)). The solid angle coverage is completed with forward copper/LAr and tungsten/LAr calorimeter modules optimised for electromagnetic and hadronic measurements, respectively.  Surrounding the calorimeters is a muon spectrometer that consists of three air-core superconducting toroidal magnets and tracking chambers, providing precision tracking for muons with  \(|\eta| < 2.7\) and trigger capability for \(|\eta| < 2.4\).\\
\newline
A two-level trigger system is used to select events for offline analysis~\cite{TRIG-2016-01}. Interesting events are selected by the first-level trigger system implemented with custom electronics which uses a subset of the detector information. This is followed by selections made by algorithms implemented in a software-based high-level trigger. The first-level trigger accepts events from the \SI{40}{\MHz} bunch crossings at a rate below \SI{100}{kHz}, which the high-level trigger further reduces in order to record events to disk at about \SI{1}{\kHz}.
\section{Observable definitions and measurement strategy}
\label{sec:strategy}
This paper presents measurements for six event-shape variables using hadronic jets. For each event, the thrust axis $\hat{n}_{\mathrm{T}}$ is defined as the direction with respect to which the projection of the jet momenta is maximised \cite{Brandt64, Farhi77}. The transverse thrust $T_{\perp}$ and its minor component $T_\text{m}$ can be expressed with respect to $\hat{n}_{\mathrm{T}}$ as
\begin{equation}
T_{\perp} = \frac{\sum_i \left|\vec{p}_{\mathrm{T},i} \cdot \hat{n}_{\mathrm{T}}\right|}{\sum_i |\vec{p}_{\mathrm{T},i}|}; \ \ \ \ \ T_\text{m} = \frac{\sum_i \left|\vec{p}_{\mathrm{T},i} \times \hat{n}_{\mathrm{T}}\right|}{\sum_i |\vec{p}_{\mathrm{T},i}|},
\label{eq:tauT}
\end{equation}
where the index $i$ runs over all jets in the event. It is also useful to define $\tau_{\perp} = 1-T_{\perp}$, so lower values of $\tau_{\perp}$ indicate a back-to-back, dijet-like configuration. The range of allowed values for these variables is, by construction, $0 \leq \tau_{\perp} < 1-2/\pi$ and $0 \leq T_\text{m} < 2/\pi$. Higher values of $\tau_\perp$ indicate a larger energy flow orthogonal to the thrust axis, while large values of~$T_\text{m}$ indicate a large energy flow outside the plane spanned by the thrust and the beam axes.\\
\newline
The sphericity $S$ and aplanarity $A$ encode information on the isotropy of the final-state energy distribution. These two observables are defined in terms of the eigenvalues of the linearised sphericity tensor of the event \cite{Par78, Don79}, given by
\begin{equation}
\mathcal{M}_{xyz} = \frac{1}{\sum_i |\vec{p}_i|} \sum_i\frac{1}{|\vec{p}_i|} \left(
\begin{array}{ccc}
p_{\mathrm{x},i}^2 & p_{\mathrm{x},i}p_{\mathrm{y},i} & p_{\mathrm{x},i}p_{\mathrm{z},i}\\
p_{\mathrm{y},i}p_{\mathrm{x},i} & p_{\mathrm{y},i}^2 & p_{\mathrm{y},i}p_{\mathrm{z},i}\\
p_{\mathrm{z},i}p_{\mathrm{x},i} & p_{\mathrm{z},i}p_{\mathrm{y},i} & p_{\mathrm{z},i}^2
\end{array}
\right).
\label{eq:spTensor}
\end{equation}
 
Its eigenvalues $\{\lambda_k\}$, which satisfy $\sum_k\lambda_k = 1$ by definition, are ordered so that $\lambda_1 > \lambda_2 > \lambda_3$, and the corresponding event shapes are defined as
\begin{equation}
S = \frac{3}{2}(\lambda_2+\lambda_3); \ \ \ A = \frac{3}{2}\lambda_3.
\label{eq:linSpher}
\end{equation}
 
$S$ takes values between 0 and 1, with larger values indicating more spherical events. $A$ takes values between 0 and 1/2 and is a measure of the extent to which the radiation is contained in the plane defined by the two first eigenvectors of the sphericity tensor defined in Eq.~\ref{eq:spTensor}. The larger the value of $A$, the less planar the event.\\
\newline
The transverse linearised sphericity tensor is constructed using only the transverse momentum components:
\[
\mathcal{M}_{xy} = \frac{1}{\sum_i |\vec{p}_i|} \sum_i\frac{1}{|\vec{p}_i|} \left(
\begin{array}{ccc}
p_{\mathrm{x},i}^2 & p_{\mathrm{x},i}p_{\mathrm{y},i}\\
p_{\mathrm{y},i}p_{\mathrm{x},i} & p_{\mathrm{y},i}^2
\end{array}
\right).
\]
 
\begin{figure}[H]
\centering
\vspace{0.5cm}
\includegraphics[width=0.49\textwidth]{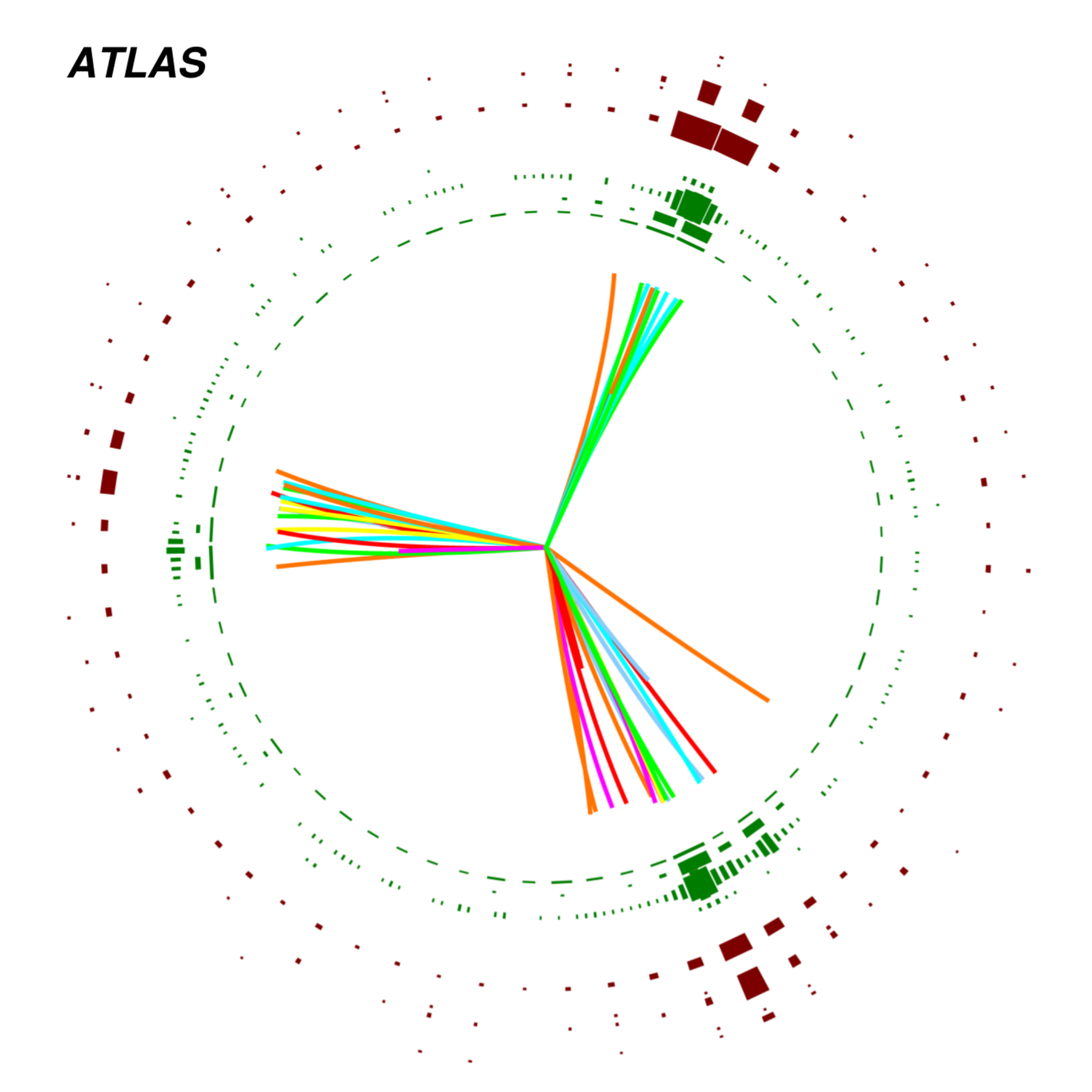}
\includegraphics[width=0.49\textwidth]{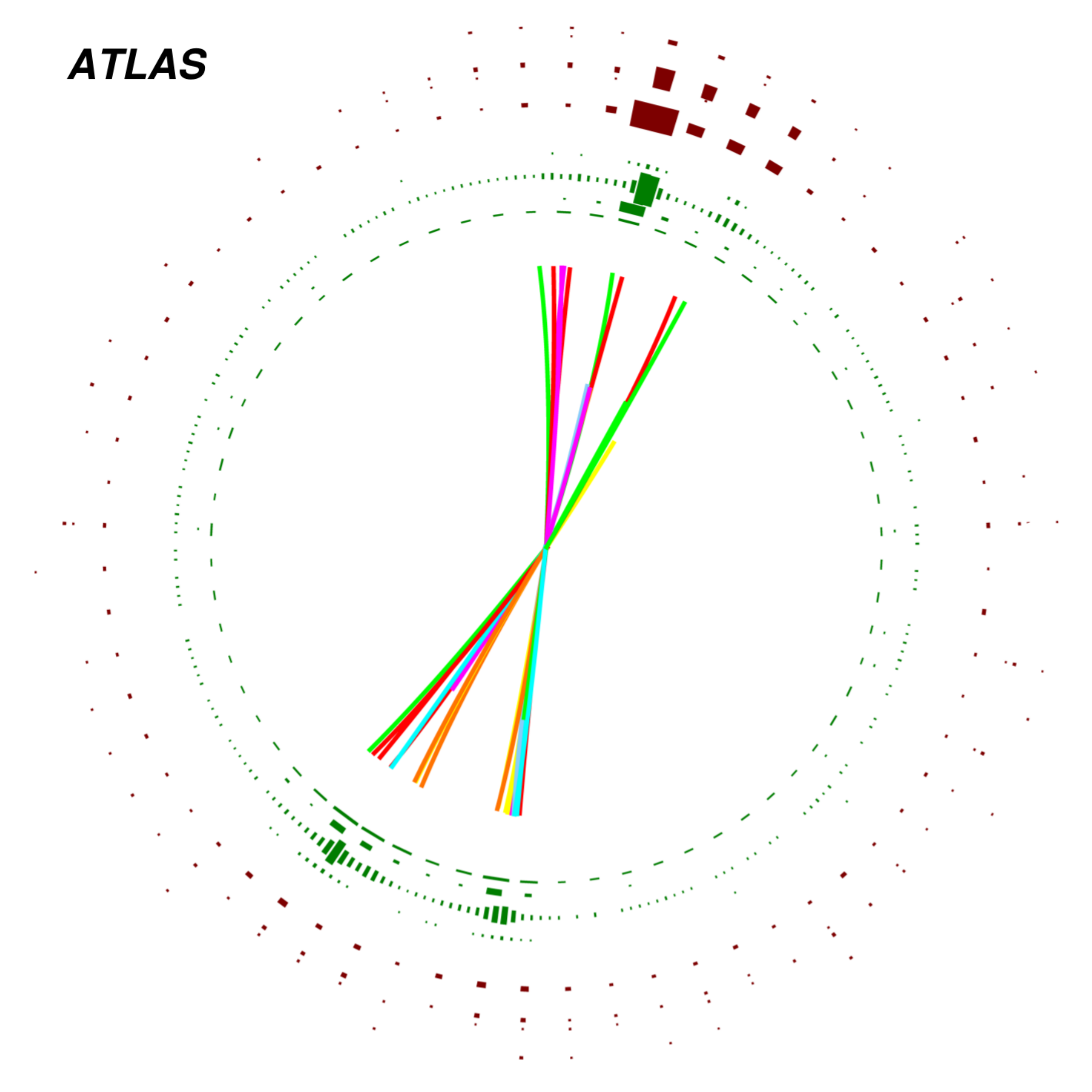}\\
\caption{Transverse plane projection of a three-jet event with high values of $\tau_{\perp}$ and $S_{\perp}$ (left), and a five-jet event with low values $\tau_{\perp}$ and $S_\perp$ (right). The colours are chosen for illustrative purposes.}
\label{fig:eventDisplay}
\end{figure}

Its eigenvalues $\{\mu_k\}$, which satisfy $\sum_k\mu_k = 1$ by definition, are ordered so that $\mu_1 > \mu_2$ and the corresponding transverse sphericity event shape is defined as
 
\begin{equation}
S_{\perp} = \frac{2\mu_2}{\mu_1+\mu_2}.
\label{eq:ST}
\end{equation}
It takes values between 0 and 1, with large (small) values indicating isotropic (back-to-back) events in the transverse plane.\\
\newline
To illustrate the meaning of the event-shape variables, Figure~\ref{fig:eventDisplay} shows two different multijet final states. The first represents a three-jet event with large values of $\tau_\perp$ and $S_\perp$. The second represents a five-jet event with low values of $\tau_\perp$ and $S_\perp$.
 
The quantities in Eq. \ref{eq:linSpher} correspond to linear combinations of the eigenvalues of the sphericity tensor. However, one may consider quadratic and cubic combinations of the eigenvalues $\{\lambda_i\}$ \cite{Ell81}, such as
\begin{gather}
C = 3(\lambda_1\lambda_2 + \lambda_1\lambda_3 + \lambda_2\lambda_3)\label{eq:C}, \\
D = 27(\lambda_1\lambda_2\lambda_3).\label{eq:D}
\end{gather}
The quantities defined in Eqs.~\ref{eq:C} and \ref{eq:D} are restricted to the range $[0,1]$. These are also useful observables to characterise multijet events. Since $C$ is defined by products of eigenvalue pairs, it vanishes for two-jet events, while $D$, which is defined by multiplying the three eigenvalues, vanishes for events in which all jet momenta lie on the same plane.\\
\newline
To study the dependence of the observables on the event topology and energy scale, each of the six event-shape observables is measured as a function of $n^{\textrm{jet}}$ and $H_{\mathrm{T}2}$. Events that satisfy the selection requirements are classified in bins of $n^{\textrm{jet}}$ ($=2$, $3$, $4$, $5$ and $\geq 6$) and $H_{\mathrm{T}2}$ (1 TeV $<H_{\mathrm{T}2}<$ 1.5 TeV, 1.5~TeV~$<H_{\mathrm{T}2}<$~2.0~TeV, $H_{\mathrm{T}2}>$~2~TeV).\\
\newline
A measurement of the multijet production cross section in the different $H_{\mathrm{T}2}$ bins is performed in the same fiducial phase space in which the event-shape observables are measured, i.e.\ in events with $2, 3, 4, 5$ or $\ge 6$ jets. Since many of the experimental uncertainties that affect the measurement of the event-shape observables are correlated between $n^\textrm{jet}$ bins, these measurements  are presented normalised to the inclusive dijet cross section in bins of $H_{\mathrm{T}2}$. In this way, the experimental uncertainties discussed in Section~\ref{sec:syst_uncert} are significantly reduced while preserving important physics information, such as the relative shape of the distributions.
 
\section{Data and Monte Carlo samples}
\label{sec:MC_simulation}
The dataset used in this analysis comprises the data taken from 2015 to 2018 at a centre-of-mass energy of $\sqrt{s} = 13~\TeV$. After applying quality criteria to ensure good ATLAS detector operation, the total integrated luminosity useful for data analysis is 139~$\ifb$. The average number of inelastic $pp$ interactions produced per bunch crossing for the dataset considered, hereafter referred to as `pile-up', is $\langle\mu\rangle = 33.6$.\\
\newline
Several MC samples were used for this analysis; they differ in the matrix element (ME) calculation and/or the parton shower (PS). In order to populate all regions of the spectra, these samples are divided into subsamples with differing kinematic characteristics. The samples were produced using the \textsc{Pythia} \cite{Sjostrand:2006za, Sjostrand:2014zea}, \textsc{Sherpa} \cite{sherpa}, \textsc{Herwig} \cite{Bahr:2008pv,Bellm:2015jjp,Bellm:2017bvx} and \textsc{MadGraph5}\_a\textsc{MC@NLO} (hereafter referred to as \textsc{MG5}\_a\textsc{MC})~\cite{madgraph} generators.\\
\newline
The \textsc{Pythia} sample was generated using \textsc{Pythia 8.235}. The matrix element (ME) was calculated for the $2\to 2$ process. The parton shower algorithm includes initial- and final-state radiation based on the dipole-style $\pt$-ordered evolution, including $\gamma\to q\bar{q}$ branchings and a detailed treatment of the colour connections between partons \cite{Sjostrand:2006za}. The renormalisation and factorisation scales were set to the geometric mean of the squared transverse masses of the two outgoing particles (labelled 3 and 4), i.e. $\sqrt{m^2_{\mathrm{T_3}} \cdot m^2_{\mathrm{T_4}}}$ = $\sqrt{(\pt^2 + m_3^2) \cdot (\pt^2 + m_4^2)}$. The \textsc{NNPDF 2.3 LO} PDF set~\cite{nnpdf} was used in the ME generation, in the parton shower, and in the simulation of multi-parton interactions (MPI). The ATLAS A14~\cite{ATL-PHYS-PUB-2014-021} set of tuned parameters (tune) is used for the parton shower and MPI, whilst hadronisation was modelled using the Lund string model~\cite{Andersson:1983ia,Sjostrand:1984ic}. The \textsc{Pythia} sample contains per-event weights that allow the estimation of uncertainties due to the parton shower parameters, including the variations of the renormalisation scale for QCD initial- and final-state radiation (ISR -- FSR) and variations of the non-singular terms of the splitting functions.\\
\newline
The \textsc{Sherpa} sample was generated using \textsc{Sherpa 2.1.1}. The ME calculation is included for the $2\to 2$ and $2\to 3$ processes at leading order (LO), and the \textsc{Sherpa} parton shower~\cite{Catani:2001cc, Krauss:2002up} with $\pt$ ordering was used for the showering. Matrix element renormalisation and factorisation scales for $2\to 2$ processes were set to the harmonic mean of the Mandelstam variables $s$, $t$ and $u$~\cite{mandelstam}, whereas the Catani--Marchesini--Webber (CMW)~\cite{CMWscale} scale was chosen for the additional emission in $2\to 3$ processes. The \textsc{CT14 NNLO}~\cite{ct14} PDF set was used for the matrix element calculation, while the parameters used for the modelling of the MPI and the parton shower were set according to the \textsc{CT10} tune~\cite{ct10}. The \textsc{Sherpa} sample makes use of the dedicated \textsc{Sherpa} AHADIC model for hadronisation \cite{Winter:2003tt}, which is based on the cluster fragmentation algorithm~\cite{Webber:1983if}.\\
\newline
The \textsc{MG5}\_a\textsc{MC}+\textsc{Pythia~8} sample was generated using \textsc{MadGraph5}\_a\textsc{MC@NLO 2.3.3}. The  calculation includes matrix elements computed at leading order for up to four final-state partons, using the \textsc{NNPDF 3.0 NLO}~\cite{nnpdf3} PDF set, and merged with the CKKW-L prescription~\cite{Lonnblad:2001iq,Lonnblad:2011xx}. The renormalisation and factorisation scales were set to the transverse mass of the $2\to2$ system that results from the $k_t$ clustering~\cite{ktAlgorithm} and the  merging scale was set to 30~\GeV. The parton shower and hadronisation were handled by \textsc{Pythia~8.212}. The ATLAS A14 tune with the \textsc{NNPDF 2.3 LO} PDF set was used for the shower and MPI, and the Lund string model was used for the modelling of hadronisation.\\
\newline
Finally, two \textsc{Herwig} samples were generated using \textsc{Herwig 7.1.3} at next-to-leading order. This includes NLO accuracy for the $2\to 2$ process and LO accuracy for the $2\to 3$ process. The ME was calculated using \textsc{Matchbox} \cite{Platzer:2011bc} with the \textsc{MMHT2014 NLO} PDF~\cite{mmht}. The renormalisation and factorisation scales were set to the $\pt$ of the leading jet. The first sample uses an angle-ordered parton shower, while the second sample uses a dipole-based parton shower. In both cases, the parton shower was interfaced to the ME calculation using the \textsc{MC@NLO} matching scheme. The angle-ordered shower evolves on the basis of 1$\to $2  splittings  with  massive  DGLAP  functions using  a  generalised  angular  variable  and  employs  a  global recoil scheme once showering has terminated. The dipole-based shower uses 2$\to $3 splittings with Catani--Seymour kernels with an ordering in transverse momentum and so is able to perform recoils on an emission-by-emission basis. For both \textsc{Herwig} samples, the parameters that control the MPI and parton shower simulation were set according to the \textsc{H7-UE-MMHT} tune~\cite{mmht}, and the hadronisation was modelled by means of the cluster fragmentation algorithm.\\
\newline
The main features of the samples described above are summarised in Table~\ref{tab:mcSamples}.
\begin{table}[H]
\small
\caption{Properties of the Monte Carlo samples used in the analysis, including the perturbative order in $\alpha_\text{s}$, the number of final-state partons, the PDF set, the parton shower algorithm, the renormalisation and factorisation scales and the value of $\alpha_\text{s}(m_\Zboson)$ for the matrix element.}
\begin{center}
\begin{tabular}{c|c|c|c|c|c|c}
\T\B Generator & ME order & FS partons & PDF set & Parton shower & Scales $\mu_\text{R},\mu_\text{F}$ & $\alpha_\text{s}(m_\Zboson)$\\
\hline
\hline
\T\T\B \textsc{Pythia} & LO & 2 & \textsc{NNPDF 2.3 LO} & $\pt$-ordered & $(m_{\mathrm{T}3}\cdot m_{\mathrm{T}4})^{\frac{1}{2}} $ & 0.140\\
\hline
\multirow{2}{*}{\T\B \textsc{Sherpa}} & \multirow{2}{*}{LO} & \multirow{2}{*}{2,3} & \multirow{2}{*}{\textsc{CT14 NNLO}} & \multirow{2}{*}{CSS (dipole)} & \T\B $H(s,t,u)$ [$2\to 2$] & \multirow{2}{*}{0.118}\\
&    &     &                    &              & CMW [$2\to 3$]                       & \\
\hline
\T\B \textsc{MG5}\_a\textsc{MC} & LO & 2,3,4 & \textsc{NNPDF 3.0 NLO} & $\pt$-ordered & $m_{\mathrm{T}}$ & 0.118\\
\hline
\multirow{2}{*}{\T\B \textsc{Herwig}} & \multirow{2}{*}{NLO} & \multirow{2}{*}{2,3} & \multirow{2}{*}{\textsc{MMHT2014 NLO}} & \T\B Angle-ordered & \multirow{2}{*}{$\max_{i}\left\{p_{\mathrm{T}i}\right\}_{i=1}^N$} & \multirow{2}{*}{0.120}\\
&                                    &  & & Dipole &\\
\end{tabular}
\end{center}
\label{tab:mcSamples}
\end{table}
 
The \textsc{Pythia}, \textsc{Sherpa} and \textsc{Herwig} samples were passed through the $\GEANT4$-based~\cite{Agostinelli:2002hh} ATLAS detector-simulation program~\cite{SOFT-2010-01} since they were also used to unfold the measurements to the particle level, as described in Section~\ref{sec:unfolding}. They are reconstructed and analysed with the same processing chain as the data. The \textsc{MG5}\_a\textsc{MC} samples are used for comparison at particle level.\\
\newline
The generation of the simulated event samples includes the effect of multiple $pp$ interactions per bunch crossing, as well as the effect on the detector response of interactions from bunch crossings before or after the one containing the hard interaction. In addition, during the data-taking, some modules (so-called ``dead-tile modules'') situated in various $\eta$-$\phi$ regions of the ATLAS hadronic calorimeter were found to be malfunctioning for some periods of time, leading to poorly reconstructed jets in these regions. The resulting dead regions in the hadronic calorimeter were included in the simulation for the \textsc{Pythia}, \textsc{Sherpa} and \textsc{Herwig} samples.

\section{Event selection and object reconstruction}
\label{sec:data_selection}
Events with high-$\pt$ jets are preselected using a single-jet trigger with a minimum $\pt$ threshold of 460 GeV. Events are required to have at least one reconstructed vertex that contains two or more associated tracks with transverse momentum $\pt > 500~\MeV$. The reconstructed vertex that maximises $\sum \pt^2$, where the sum is performed over tracks associated with the vertex, is chosen as the primary vertex.\\
\newline
Jets are reconstructed using the anti-$k_t$ algorithm \cite{antikt} with radius parameter $R=0.4$ using the \textsc{FastJet} program~\cite{fastjet}. The inputs to the jet algorithm are particle-flow objects~\cite{pflow}, which make use of both the calorimeter and the inner-detector information to precisely determine the momenta of the input particles. The jet calibration procedure includes energy corrections for pile-up, as well as angular corrections. Effects due to energy losses in inactive material, shower leakage, the parameterisation of the magnetic field and inefficiencies in energy clustering and jet reconstruction are taken into account. This is done using a simulation-based correction, in bins of $\eta$ and $\pt$, derived from the relation of the reconstructed jet energy to the energy of the corresponding particle-level jet, not including muons or non-interacting particles. In a final step, an in situ calibration corrects for residual differences in the jet response between the MC simulation and the data using $\pt$-balance techniques for dijet, $\gamma+$jet, $\Zboson$+jet and multijet final states.\\
\newline
The selected jets must have $\pt > 100~\GeV$ and $|\eta| < 2.4$. These requirements reject pile-up jets and reduce experimental uncertainties. In addition, jets are required to satisfy quality criteria that reject beam-induced backgrounds (jet cleaning)~\cite{cleaning}. The efficiency of this requirement for selecting good jets with $\pt >$ 100~\GeV~is larger than 99.5\%. Events are required to have at least two selected jets. The two leading jets are further required to satisfy $H_{\mathrm{T}2} > 1~\TeV$. This requirement ensures a trigger efficiency of $\approx$100\%. About 57.5 million events in data satisfy the selection criteria. Figure~\ref{fig:reco} shows the detector level distributions for the selected jets $\pt$ and $H_{\mathrm{T}2}$, along with MC predictions.
\begin{figure}[H]
\centering
\includegraphics[width=0.49\textwidth]{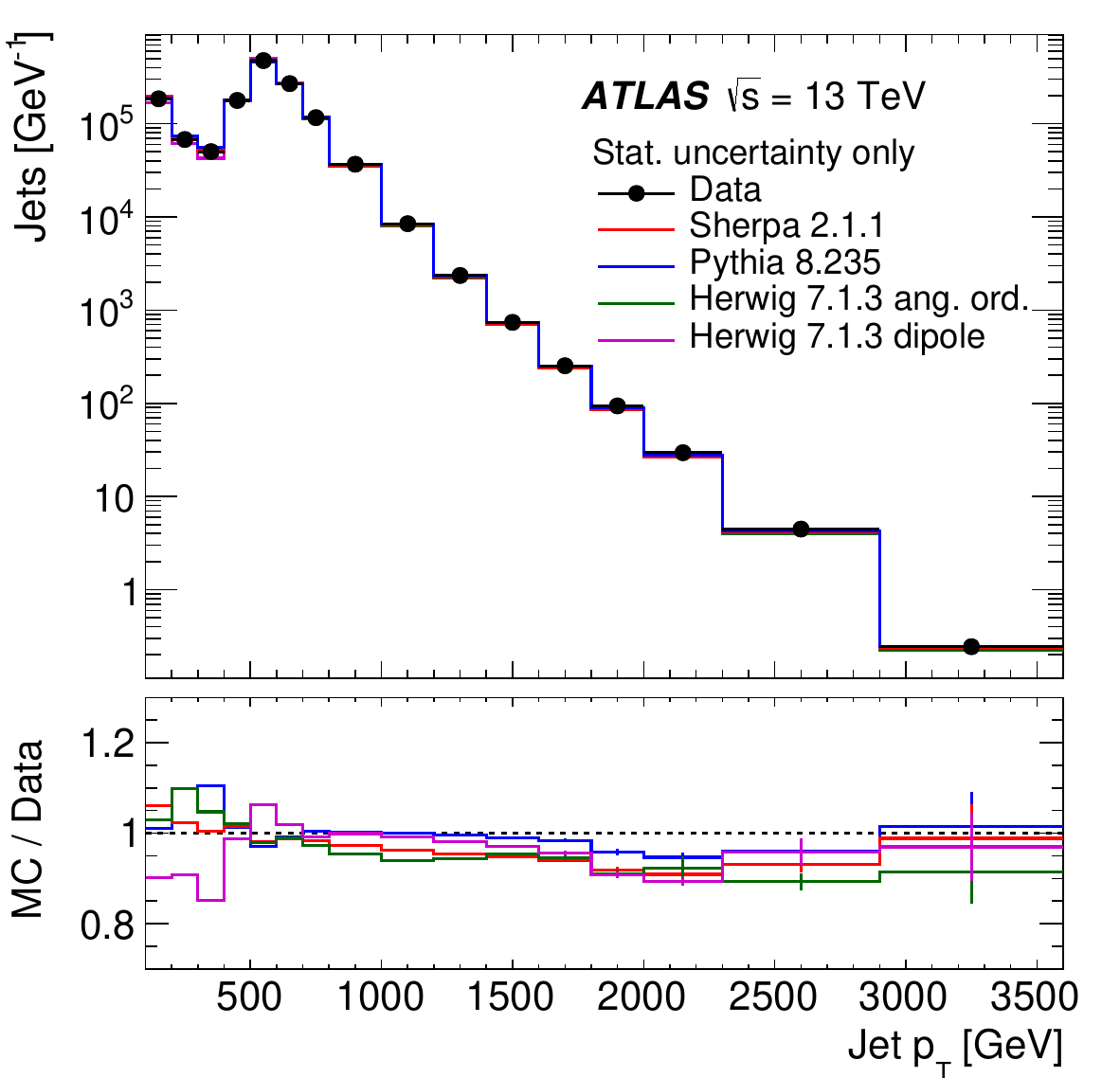}
\includegraphics[width=0.49\textwidth]{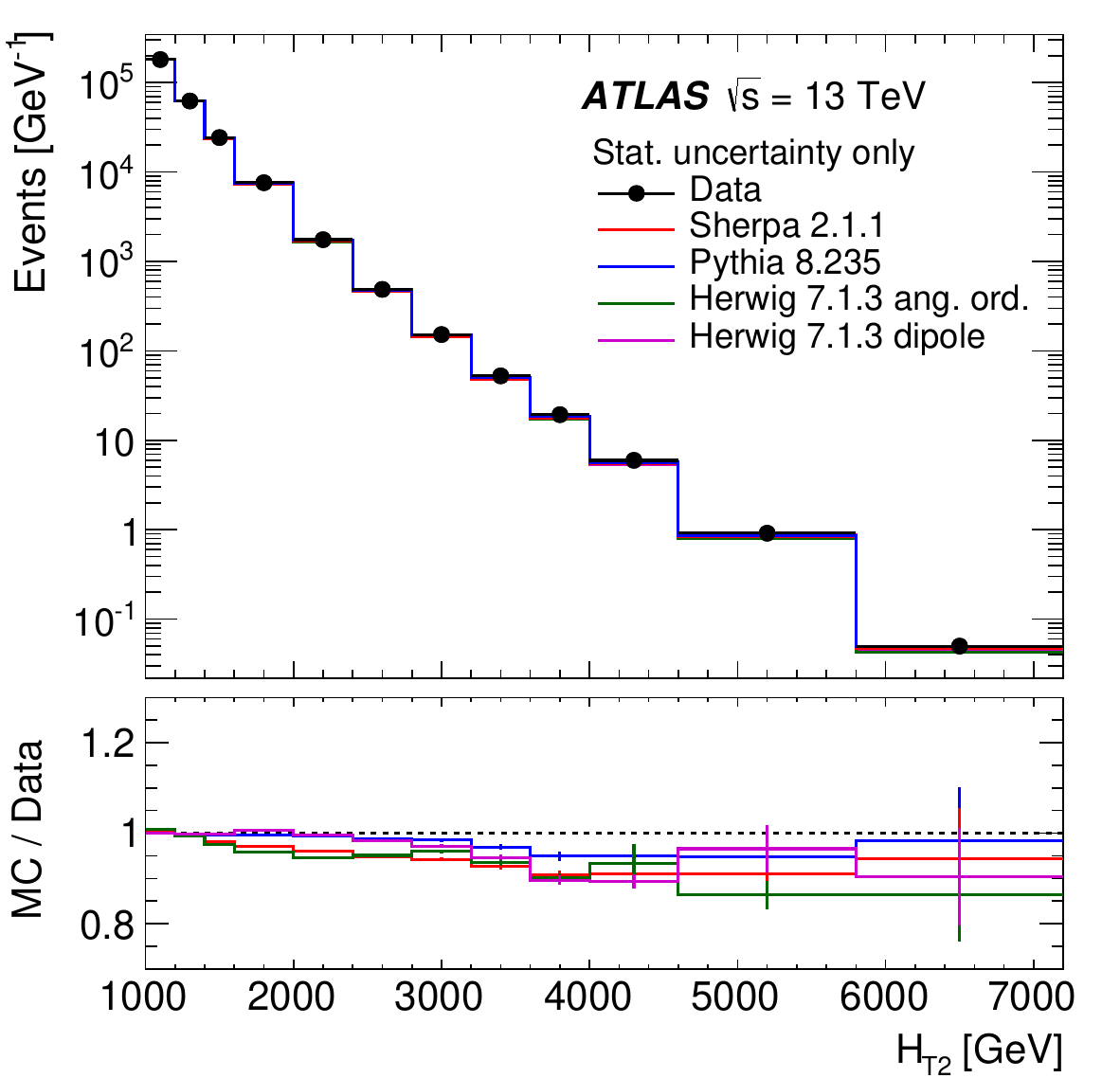}
\caption{Detector level distributions of the transverse momenta of all jets (left) and the scalar sum of transverse momenta of the two leading jets (right), together with MC predictions. Only statistical uncertainties are shown.}
\label{fig:reco}
\end{figure}
The binning of the event-shape distributions is chosen as a compromise between maximising the number of bins while minimising migration between bins due to the resolution of the measured variables.
\section{Unfolding to particle level}
\label{sec:unfolding}
In order to make meaningful comparisons with particle-level MC predictions, the event-shape distributions need to be corrected for distortions induced by the response of the ATLAS detector and associated reconstruction algorithms. The fiducial phase-space region is defined at particle level for all particles with a mean decay length $c\tau>$ 10 mm; these particles are referred to as `stable'. Particle-level jets are reconstructed using the anti-$k_t$ algorithm with $R=0.4$ using stable particles, excluding muons and neutrinos. The fiducial phase space closely follows the event selection criteria defined in Section~\ref{sec:data_selection}. Particle-level jets are required to have $\pt > 100~\GeV$ and $|y| < 2.4$, where $y$ represents the rapidity. Events with at least two particle-level jets are considered. These events are also required to have $H_{\mathrm{T}2} > 1~\TeV$.\\
 
The unfolding is performed using an iterative algorithm based on Bayes' theorem \cite{bayes}. The algorithm takes into account inefficiencies and resolution effects due to the detector response that lead to bin migrations between the detector-level and particle-level phase spaces. For each observable, the method makes use of a transfer matrix, $M_{ij}$, obtained from MC simulation, that parameterises the probability of an event generated in bin $i$ to be reconstructed in bin $j$. The correction can thus be written as a linear equation
 
\[
R_i = \sum_{j=1}^N \frac{\mathcal{E}_j}{\mathcal{P}_i} M_{ij} T_j.
\]
 
The quantities $R_i$ and $T_j$ are the contents of the detector-level distribution in bin $i$ and the contents of the particle-level distribution in bin $j$, respectively, while the factors $\mathcal{E}_j$ and $\mathcal{P}_i$ are the efficiency and the purity, which are estimated using the MC simulation. Migrations between detector- and particle-level phase spaces due to different values of $n^{\textrm{jet}}$ and $H_{\textrm{T2}}$ are taken into account in the unfolding procedure for event-shape observables. Due to the fine binning of the measured distributions, detector resolution is the primary cause of bin migrations between detector- and particle-level phase spaces, followed by the jet energy resolution, which leads to different values of $n^{\textrm{jet}}$ between the detector and particle levels. The efficiency $\mathcal{E}_j$ is used to correct for events in the particle-level phase space which are not reconstructed at detector level. The binned efficiency is defined by the number of MC events that satisfy all the selection requirements at both the detector and particle levels and are generated and reconstructed in the same event-shape bin, with the same $n^{\textrm{jet}}$ and $H_{\textrm{T2}}$ range, divided by the number of events generated in the same event-shape, $n^{\textrm{jet}}$ and $H_{\textrm{T2}}$ bin. For low values of the event-shape variables the efficiency is typically close to 80\%, while for large values it decreases to $\approx$ 40\%. In addition, the binned efficiency tends to have lower values at higher $n^{\textrm{jet}}$ for the same event-shape bin. The purity $\mathcal{P}_i$ is used to correct for events in the detector-level phase space that do not have a particle-level counterpart. The binned purity is defined as the number of events that satisfy selection requirements at particle and detector levels divided by those reconstructed in the same event-shape, $n^{\textrm{jet}}$ and $H_{\mathrm{T}2}$ bin. Similarly to the bin-by-bin efficiency, the purity is close to 80\% for low values of the event-shape variables and decreases to 40--50\% for large values.  The bin-by-bin purity tends to have lower values for the same event-shape bin at higher $n^{\textrm{jet}}$. Since the bin-by-bin purity and efficiency distributions are similar, and the migrations in the transfer matrices are mainly between neighbouring bins of the event-shape distributions, the impact of the unfolding on the shape of the distributions is modest.
The results obtained by unfolding the data with \textsc{Pythia} are used to obtain the nominal differential cross sections, whereas \textsc{Sherpa} and \textsc{Herwig} MC predictions are used to estimate the systematic uncertainty due to the model dependence, as discussed in Section~\ref{sec:syst_uncert}.
\section{Experimental uncertainties}
\label{sec:syst_uncert}
The dominant sources of systematic uncertainty in the measurements arise from imperfect knowledge of the jet energy scale and resolution and the modelling of the strong interaction. The systematic uncertainties are propagated through the unfolding via their impact on the transfer matrices.
The inclusive cross section for events with at least two jets is recomputed for each systematic variation and used consistently in the analysis. As an example, the breakdown of the relative systematic uncertainties in the measurement of the normalised differential cross sections as functions of $\tau_{\perp}$ and $A$ is shown in Figure~\ref{fig:syst}.
 
\begin{figure}[H]
\centering
\includegraphics[scale=0.38]{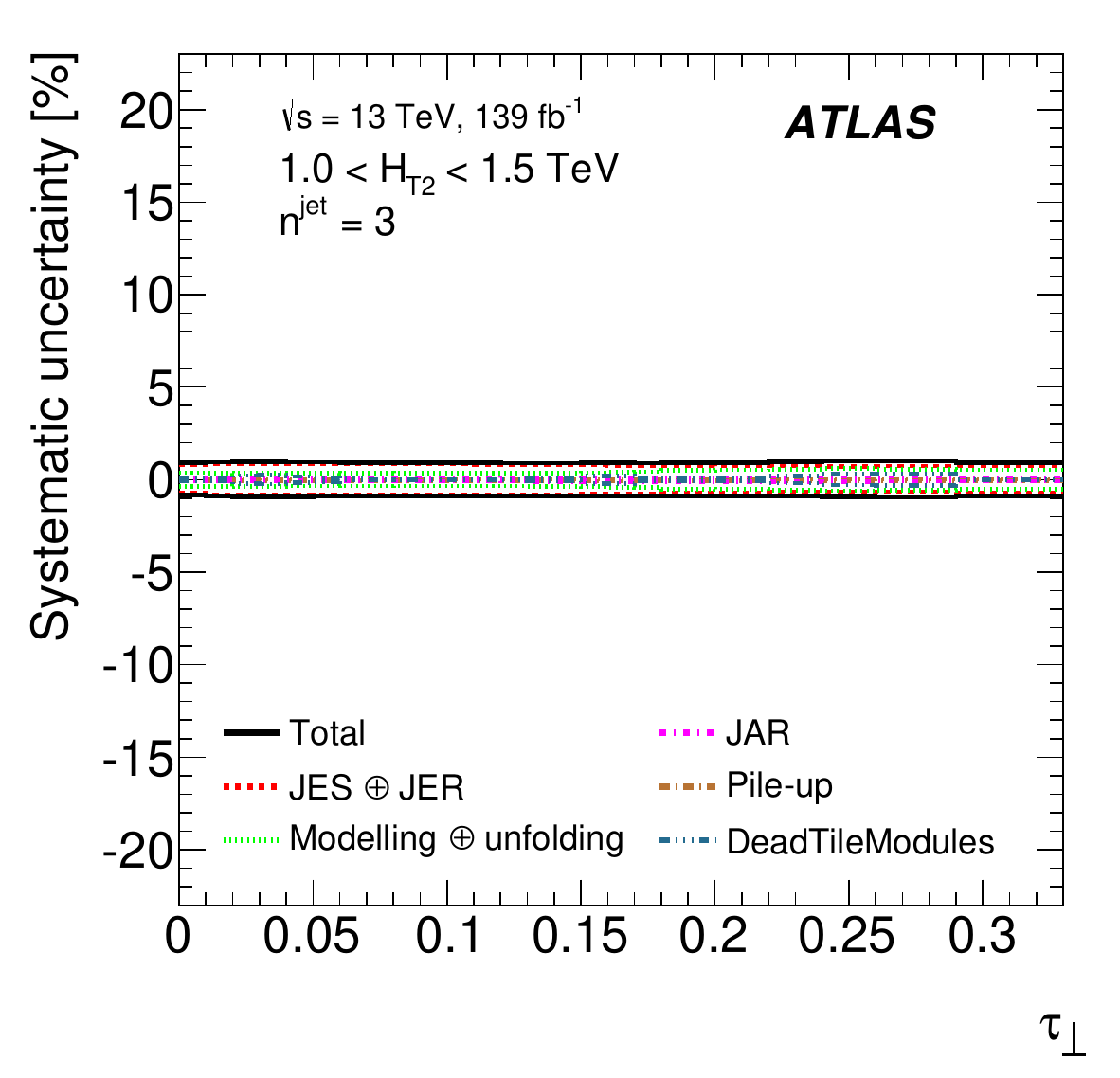}
\includegraphics[scale=0.38]{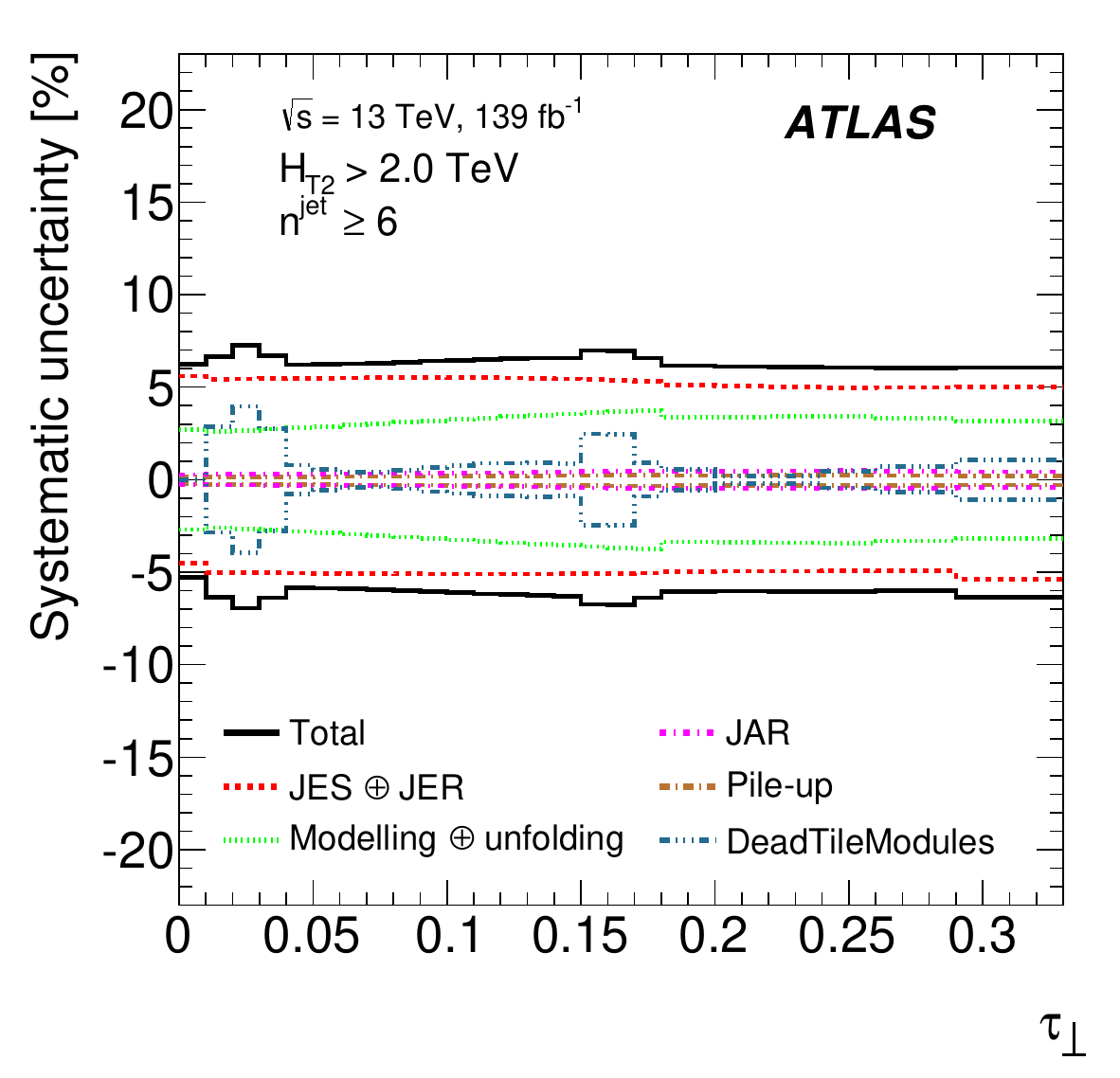}\\
 
\includegraphics[scale=0.38]{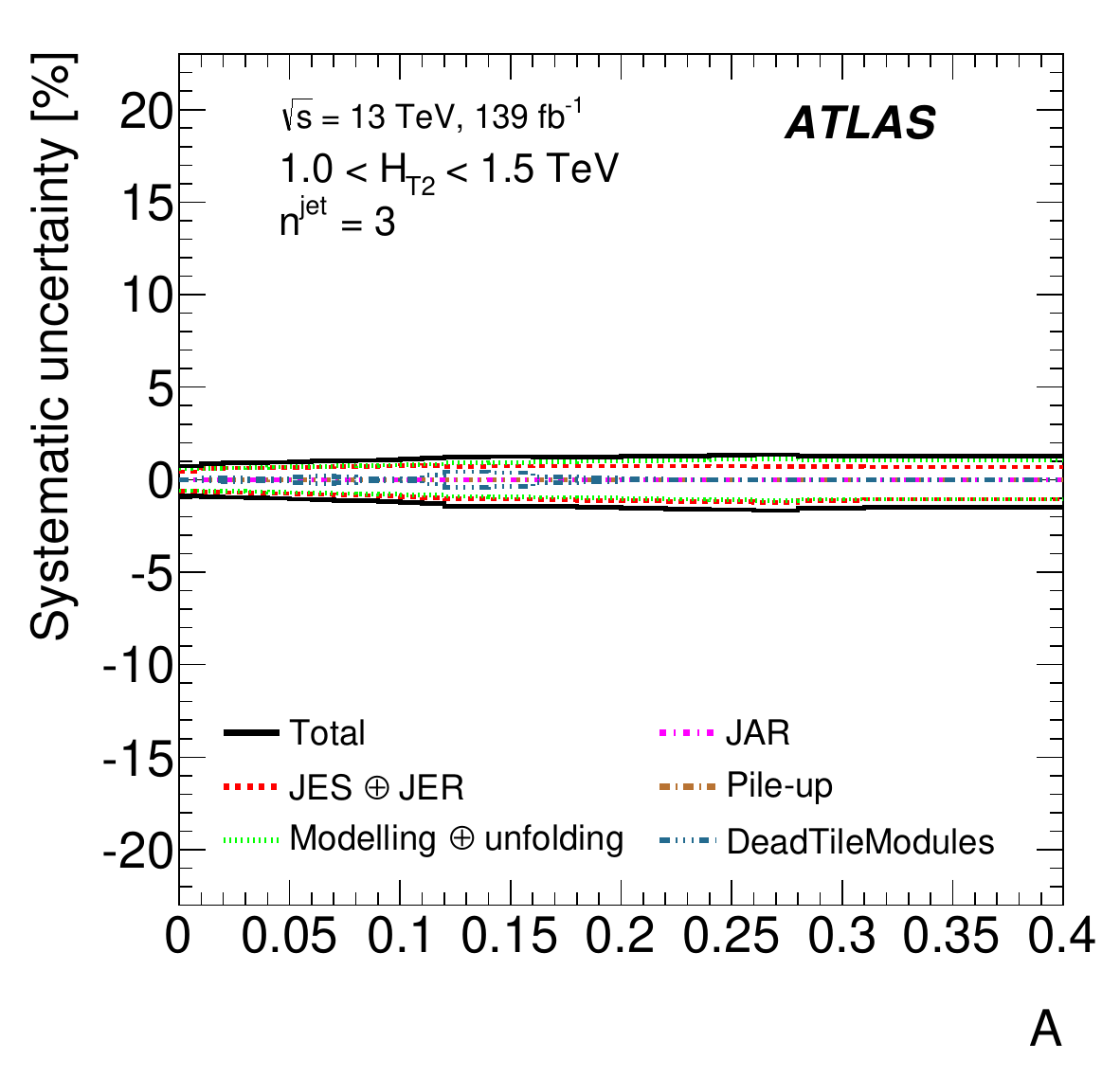}
\includegraphics[scale=0.38]{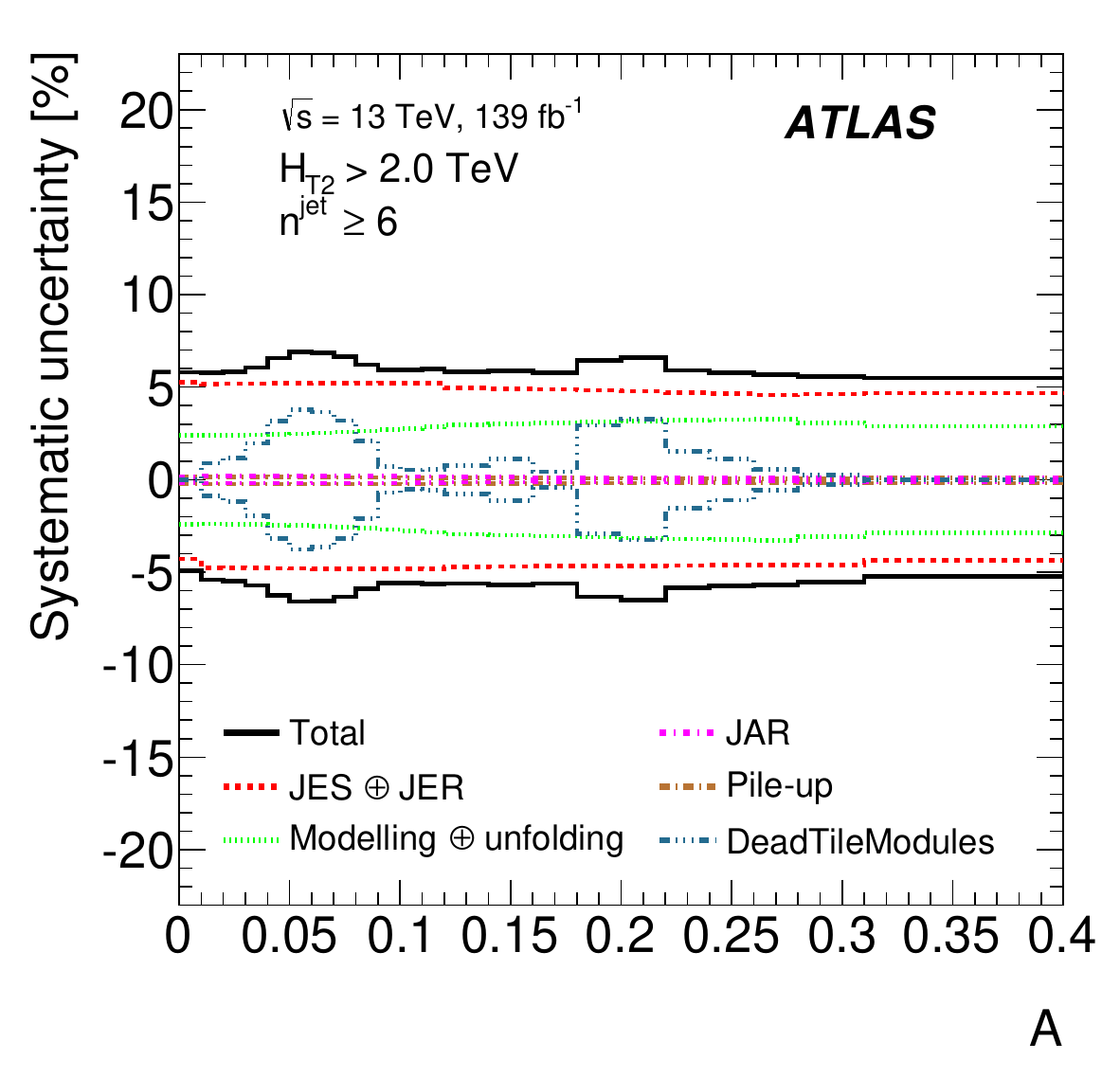}
\caption{Breakdown of the systematic uncertainties as a function of $\tau_\perp$ (top) and $A$ (bottom) for selected regions of $H_{\mathrm{T}2}$ and $n^{\textrm{jet}}$.}
\label{fig:syst}
\end{figure}
 
A detailed description of the systematic uncertainties and their values for normalised event-shape observables is given below:
\begin{itemize}
\item{} Jet Energy Scale and Resolution: The jet energy scale (JES) and jet energy resolution (JER) uncertainties are estimated as described in Ref.~\cite{JES}. The JES is calibrated on the basis of the simulation, including in situ corrections obtained from data. The JES uncertainties are estimated using a decorrelation scheme comprising a set of 44 independent components, which depend on the jet $\pt$ and $\eta$. The total JES uncertainty in the $\pt$ value of individual jets is $<$ 2\% at $\pt$~=~100~GeV with a mild dependence on
$\eta$. The JER uncertainty is estimated using a decorrelation scheme involving 26 independent components. The effect of the total JER uncertainty is evaluated by smearing the energy of the jets in the MC simulation by about 1.5\% at $\pt$ = 100 GeV to about 0.5\% for $\pt$ of several hundred GeV. In this measurement, the JES and JER uncertainties are propagated by varying the energy and $\pt$ of each jet by one standard deviation of each of the independent components. The total uncertainty in the normalised event-shape distributions varies from 1\% in the lowest $n^{\textrm{jet}}$ bins to 7\% for the highest $n^{\textrm{jet}}$, while it ranges from 7\% to 14\% on the fiducial cross sections. This is the dominant source of uncertainty for high $n^{\textrm{jet}}$.
 
\item{} Jet Angular Resolution: The jet angular resolution (JAR) uncertainty is estimated conservatively by smearing the angular coordinates $(\eta,\phi)$ of the jets by the resolution in MC simulation. The $\eta$ and $\phi$ variations are done with the $p_{\textrm{T}}$ component of the jets held constant. The value of the JAR uncertainty is below 0.5\% for the normalised event-shape distributions in all regions of the phase space, while it ranges from 1\% to 2\% for the fiducial cross sections, as $n^{\textrm{jet}}$ increases.
 
\item{} Pile-up: The uncertainty from pile-up is evaluated by varying the pile-up reweighting procedure (see Section~\ref{sec:MC_simulation}) to cover the difference between the predicted inelastic cross section and the measured value~\cite{inelasticXS}. The impact of this uncertainty is below 0.5\% in all regions of the phase space, for both the event-shape and fiducial cross-section measurements. As a cross-check, the event-shape distributions were compared in different slices of $\langle\mu\rangle$ and between different data-taking periods, yielding compatible results.
 
\item{} Unfolding: The mismodelling of the data in the MC simulation is accounted for as an additional source of uncertainty. This is assessed by reweighting the particle-level distributions so that the detector-level event shapes predicted by the MC samples match those in the data. The modified detector-level distributions are then unfolded using the method described in Section~\ref{sec:unfolding}. The difference between the modified particle-level distribution and the nominal one is taken as the uncertainty. This uncertainty ranges from 0.2\% to a few per cent with increasing $n^{\textrm{jet}}$, depending on the observable under study.
 
\item{} Modelling: The modelling uncertainty is estimated by comparing the unfolded distributions using \textsc{Pythia}, \textsc{Sherpa} and \textsc{Herwig}. In order to not double count the effect of having different priors in the unfolding procedure (the so-called unfolding uncertainty), \textsc{Pythia}, \textsc{Sherpa} and \textsc{Herwig} MC predictions are weighted to describe the data. These weighted MC samples are then used to perform the unfolding and the envelope of the differences between the estimated cross sections defines the systematic uncertainty. The value of this uncertainty for the normalised event-shape measurement increases with $n^{\textrm{jet}}$ and $H_{\mathrm{T}2}$ and varies between 1\% and 4--5\%, depending on the observable under study. For the fiducial cross-section measurement this uncertainty is below 5\%.
 
\item{} Luminosity: The uncertainty in the combined 2015--2018 integrated luminosity is 1.7\% \cite{ATLAS-CONF-2019-021}, obtained using the LUCID-2 detector \cite{LUCID2} for the primary luminosity measurements. The measurements of event-shape observables are unaffected by this uncertainty, given that they are normalised to the inclusive dijet cross section, thus cancelling out the contribution of the luminosity.
 
\item{} Dead-tile modules: A systematic uncertainty is derived to address the possible bias on the measurements due to the residual mismodeling of disabled portions of the tile calorimeter by the MC simulations. New differential cross-sections are derived by vetoing events in data and MC simulation where at least one of these non-operating modules is found within the selected jets. The difference between this result and the nominal one is taken as the uncertainty. The value of this uncertainty is below 1\% in most regions of the phase space, although it can reach values up to 4\% in some regions for the highest jet multiplicity bin.
 
\end{itemize}
 
The total systematic uncertainty is estimated by adding in quadrature the effects previously listed. In addition, the statistical uncertainty of the data and MC simulation is propagated to the differential cross sections through the unfolding procedure using pseudo-experiments in order to properly take into account the statistical correlations between bins of the event-shape variables, $n^{\textrm{jet}}$ and $H_{\textrm{T2}}$ ranges. Moreover, the pseudo-experiments are also used to estimate the statistical component of each systematic uncertainty. This statistical component is reduced using the Gaussian Kernel smoothing technique~\cite{PERF-2012-01}. The values provided above are quoted after application of this procedure.\\
\newline
In general the modelling uncertainty tends to dominate at low values of $n^{\textrm{jet}}$ while, at high $n^{\textrm{jet}}$, the JES uncertainty dominates. The total systematic uncertainty is typically constant for the measured differential cross sections, but increases as a function of $n^{\textrm{jet}}$ from $\sim$1\% to $\sim$6\%. For the fiducial cross sections, the uncertainties increase from $\sim$5\% to $\sim$9\% as a function of $n^{\textrm{jet}}$.
\clearpage
\section{Results}
\label{sec:result}
 
The differential cross-section measurements are presented and compared with the MC predictions described in Section~\ref{sec:MC_simulation}. The cross section as a function of $n^{\textrm{jet}}$ is shown in Figure~\ref{fig:xsec_total}, while unfolded and normalised event-shape distributions are shown in Figures~\ref{fig:xsec_transThrust}--\ref{fig:xsec_D}. The full set of observables is presented in each $H_{\mathrm{T}2}$ and $n^{\textrm{jet}}$ bin in which the measurement is performed. The ratio of the MC prediction to the yield in data in each bin is also shown.\\
\newline
Figure~\ref{fig:xsec_total} shows the fiducial cross section as a function of $n^{\textrm{jet}}$ in different $H_{\textrm T2}$ ranges. The fiducial cross section is measured in the same phase-space regions as the differential measurements. The MC predictions are normalised to the measured integrated cross section in each $H_{\textrm{T2}}$ range to compare the shape of these predictions to the data. The normalisation factors for \textsc{Herwig} based on angle-ordered showers and \textsc{Sherpa} predictions increase as a function of $H_{\textrm{T2}}$, whereas a very small dependence of these factors on $H_{\textrm{T2}}$ is observed for \textsc{Herwig 7} based on dipole showers, \textsc{MG5}\_a\textsc{MC} and \textsc{Pythia} predictions. The normalisation factors for LO accuracy MC predictions such as \textsc{Pythia}, \textsc{MG5}\_a\textsc{MC} and \textsc{Sherpa} are expected to strongly depend on the MC tune. In particular, the \textsc{Pythia} prediction overestimates the inclusive dijet production cross section in the studied phase-space region by 30\%, which can be attributed to the large value of $\alpha_{\textrm S}$ included in the \textsc{Pythia} tune (see Table~\ref{tab:mcSamples}), whereas the \textsc{MG5}\_a\textsc{MC} prediction underestimates it by 35\%. The \textsc{Sherpa} prediction gives an adequate description of the measured integrated cross sections. In addition, an excellent description of the inclusive dijet production cross section is found for the \textsc{Herwig 7} prediction based on dipole showers, whereas \textsc{Herwig 7} prediction based on angle-ordered showers underestimates it, at most by 9\%. The \textsc{Pythia} prediction provides a good description of the shape of the differential cross section as a function of $n^{\textrm{jet}}$. The \textsc{Herwig 7} prediction based on angle-ordered showers gives a good description of the fiducial cross sections as a function of $n^{\textrm{jet}}$. \textsc{Sherpa} tends to overestimate the cross sections for $n^{\textrm{jet}} >$ 4. The \textsc{Herwig 7} based on dipole showers and \textsc{MG5}\_a\textsc{MC} predictions, while giving a good description of the cross section at low $n^{\textrm{jet}}$, underestimate the measurements at high $n^{\textrm{jet}}$.\\
\newline
Figure~\ref{fig:xsec_transThrust} shows the normalised cross section as a function of $\tau_{\perp}$. For low $n^{\textrm{jet}}$, the MC simulations tend to underestimate the measurement in the intermediate region of $\tau_{\perp}$, with the exception of the \textsc{Herwig 7} predictions. At high values of $\tau_{\perp}$, where the population of isotropic events is expected to be larger, all MC predictions underestimate the measurements. In particular, the largest deviation is found for the \textsc{Pythia} prediction where the high-$\pt$ third jet is less likely to be produced isotropically. The shape of the distributions tends to agree with data for larger $n^{\textrm{jet}}$. \textsc{Pythia} tends to overestimate the measurements at low values of $\tau_{\perp}$, whereas the \textsc{Herwig 7} prediction based on dipole showers highly underestimates the measurements in such region. The behaviour of $\tau_\perp$ as a function of $H_{\mathrm{T}2}$ indicates more isotropic events at low energies, with increasing alignment of jets with the thrust axis for higher energy scales. Figure~\ref{fig:xsec_transMinor} shows the normalised cross section as a function of $T_\text{m}$, with very similar conclusions.\\
\newline
Figure~\ref{fig:xsec_transSphericity} shows the normalised cross section as a function of $S_{\perp}$. In line with the observations for $\tau_\perp$ and $T_\text{m}$, the results show that, for low $n^{\textrm{jet}}$, \textsc{Pythia} and \textsc{Sherpa} simulations predict fewer isotropic events than in data, while the \textsc{Herwig 7} and \textsc{MG5}\_a\textsc{MC} predictions are closer to the measurements. In addition, while \textsc{Pythia} gives an adequate description in the intermediate region of $S_{\perp}$, it overestimates the measurements at low $S_{\perp}$. For larger $n^{\textrm{jet}}$, the description of the shape is improved in MC simulations, while a discrepancy in the normalisation is observed for different predictions.\\
\newline
Figure~\ref{fig:xsec_aplanarity} shows the normalised cross section as a function of $A$. In this case, the \textsc{Herwig 7} prediction with angle-ordered parton shower aligns with the rest of the MC simulations, predicting more planar events than data at low $n^{\textrm{jet}}$, while the \textsc{Herwig 7} prediction with the dipole parton shower predicts higher cross sections at high $A$ for low $n^{\textrm{jet}}$.\\
\newline
The normalised cross sections for the quadratic observable $C$ are shown in Figure~\ref{fig:xsec_C}. Here, larger spreads are shown in the lower tails of the distributions for low $n^{\textrm{jet}}$. \textsc{Pythia} and \textsc{Herwig 7} with the dipole-based parton shower predict a smaller cross section than data in these regions, while the other MC predictions overestimate the measurements. As with other event-shape observables, a larger spread is found in the normalisations at high $n^{\textrm{jet}}$.\\
\newline
The cubic observable $D$ is presented in Figure~\ref{fig:xsec_D}, with conclusions similar to those for $A$. The \textsc{Herwig 7} prediction with dipole-based parton shower predicts a higher cross section than data at high values of $D$ for low $n^{\textrm{jet}}$, while the other MC predictions have the opposite behaviour. For higher $n^{\textrm{jet}}$, the description of the shape becomes more similar among different MC predictions, with differences observed in the normalisation of the measurements.\\
\newline
Theoretical uncertainties on the MC predictions are not included in the discussion of the results, since the ME for the current predictions has leading order accuracy in the description of inclusive three-jet cross sections. This makes the usual variations of the renormalisation and factorisation scales not reliable for the estimation of theoretical uncertainties. To identify the phase-space regions of the event-shape variables which are more sensitive to the MC tune, variations of the parton shower parameters were examined in $\textsc{Pythia}$ as detailed in Section~\ref{sec:MC_simulation}. The effect of these variations on event-shape observables typically increases with jet multiplicity. Varying the FSR energy-scale parameter by a factor of two leads to differences of up to 40\% at high values of the event-shape variables i.e regions where the contribution of isotropic events is larger. On the other hand, varying the ISR energy-scale parameter typically leads to differences from 10 to 30\% at low values of the event shapes for high jet multiplicities. Finally, varying the non-singular terms of the splitting functions contributes a 5--10\% difference that is typically constant as a function of the event-shape variables.\\
\newline
In summary, none of the MC predictions investigated provide a good description of the data in all regions of the phase space. In general, at low $n^{\textrm{jet}}$, \textsc{Pythia} and \textsc{Sherpa} predictions underestimate the measurements at high values of the event-shape distributions, i.e. events in data follow a more isotropic distribution of the energy flow than those from these two predictions. \textsc{Pythia} shows discrepancies of up to 80\%. This shows the limited ability of parton shower models to simulate hard and wide angle radiation. The addition of 2 $\rightarrow$ 3 processes in the ME allows to improve the description of the measurements in such regions. While \textsc{Sherpa} can differ by up to 30\%, the \textsc{Herwig} and \textsc{MG5}\_a\textsc{MC} predictions show discrepancies of 10-20\%.  Moreover, both \textsc{Herwig 7} predictions overestimate the central regions of $\tau_{\perp}$ by up to 20\%, while differing in the description of the aplanarity: the angle-ordered parton shower gives rise to more planar events than in data, while the dipole-based parton shower overestimates the measurements at high values of $A$ for $n^{\textrm{jet}}$ = 3. In addition, the \textsc{Herwig 7} prediction with dipole-based parton shower underestimates the measurements at low values of the event-shape distributions, whereas the \textsc{Herwig 7} angle-ordered prediction gives a better description in these regions. Overall, the description of the measurements made by the \textsc{Herwig 7} prediction based on angle-ordered parton showers is better than the description by the dipole-based parton shower. This may be due to the fact that the \textsc{Herwig 7} dipole-based shower model is recently released, and the parameters have not been tuned as thoroughly as those of the more mature angle-ordered showers. Moreover, the \textsc{MG5}\_a\textsc{MC} prediction which includes up to four final-state partons in the ME gives the best overall description of the shape of the measurements in the studied $n^{\textrm{jet}}$ and $H_{\textrm{T2}}$ bins. This shows the importance of including in the ME beyond LO terms to describe the dynamics of high-$p_{\textrm{T}}$ multijet final states. At high $n^{\textrm{jet}}$, all MC simulations tend to give a similar prediction for the shape of the distributions. However, the normalisation of the predictions shows a large spread between different MC simulations in each bin of the jet multiplicity. The \textsc{Sherpa} prediction gives an adequate description of the normalisation for $n^{\textrm{jet}}\leq$ 4, although it overestimates the cross sections up to 30\% for high $n^{\textrm{jet}}$. The \textsc{MG5}\_a\textsc{MC} and the \textsc{Herwig 7} with dipole-based parton showers simulations predict cross sections up to 30\% lower for events with at least six jets. Finally, \textsc{Herwig 7} with angle-ordered parton shower and \textsc{Pythia} predictions give a reasonably good description of the normalisation of the differential cross sections for the studied jet multiplicity and $H_{\textrm{T2}}$ bins.\\
\newline
All measurements can be found in \textsc{Hepdata}~\cite{hepdata}, including these measurements binned in inclusive jet multiplicity.
 
\begin{figure}[H]
\centering
\vspace{-0.4cm}
\includegraphics[width=0.98\textwidth,height=0.465\textwidth]{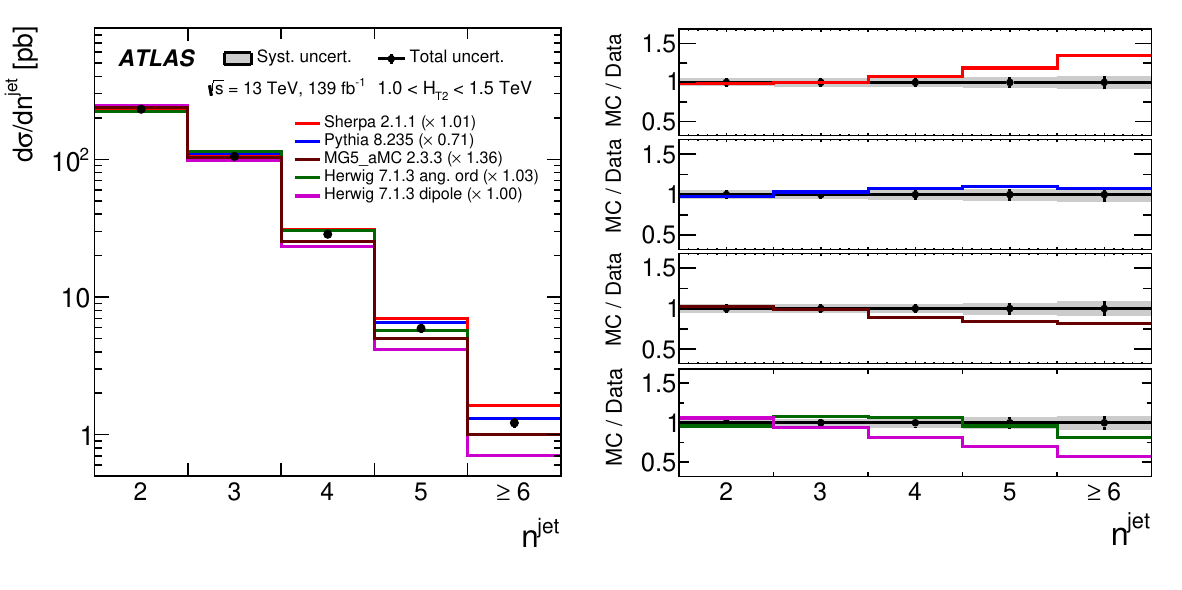}\\
\vspace{-0.5cm}
\includegraphics[width=0.98\textwidth,height=0.465\textwidth]{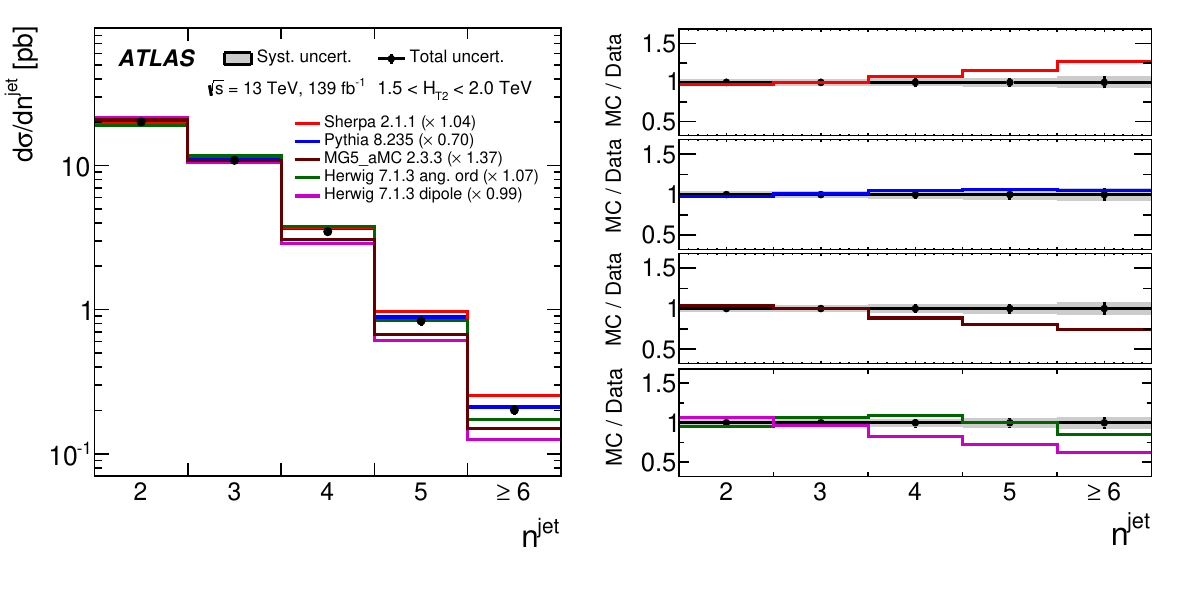}\\
\vspace{-0.5cm}
\includegraphics[width=0.98\textwidth,height=0.465\textwidth]{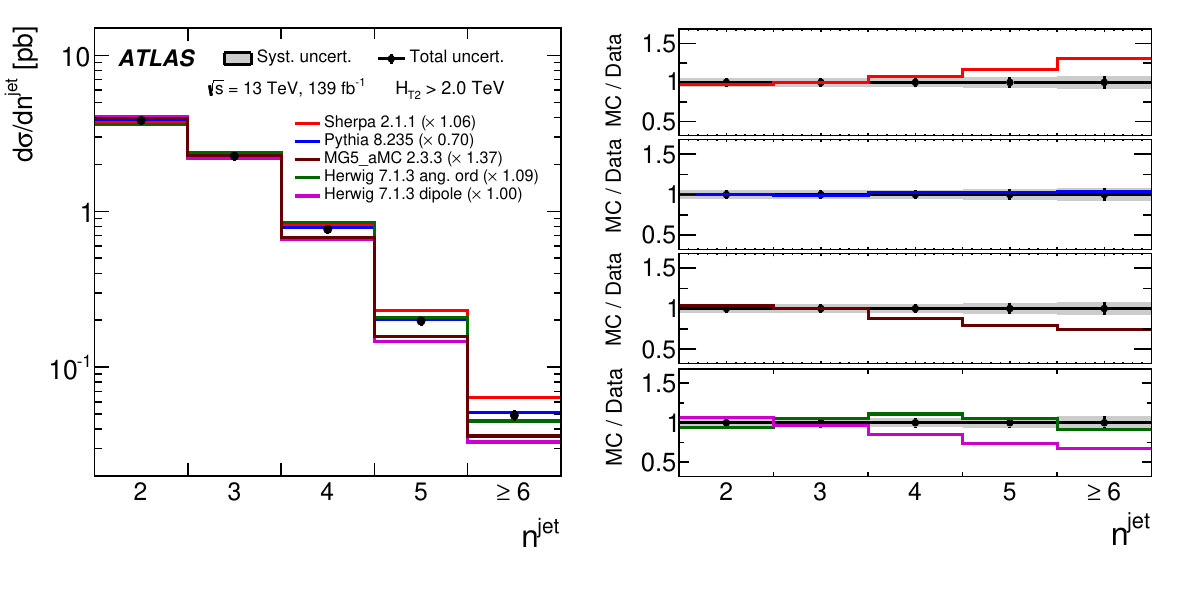}\\
\vspace{-0.55cm}
\caption{Fiducial cross section as a function of jet multiplicity. The MC predictions are normalised to the measured integrated cross section for $n^{\textrm{jet}}\geq$ 2 in each $H_{\textrm{T2}}$ bin using the factors indicated in parentheses. The right panels show the ratios of the MC distributions to the data distributions. The error bars show the total uncertainty (statistical and systematic added in quadrature) and the grey bands in the right panels show the systematic uncertainty.}
\label{fig:xsec_total}
\end{figure}
 
\begin{figure}[H]
\centering
\vspace{-0.4cm}
\includegraphics[width=0.98\textwidth,height=0.465\textwidth]{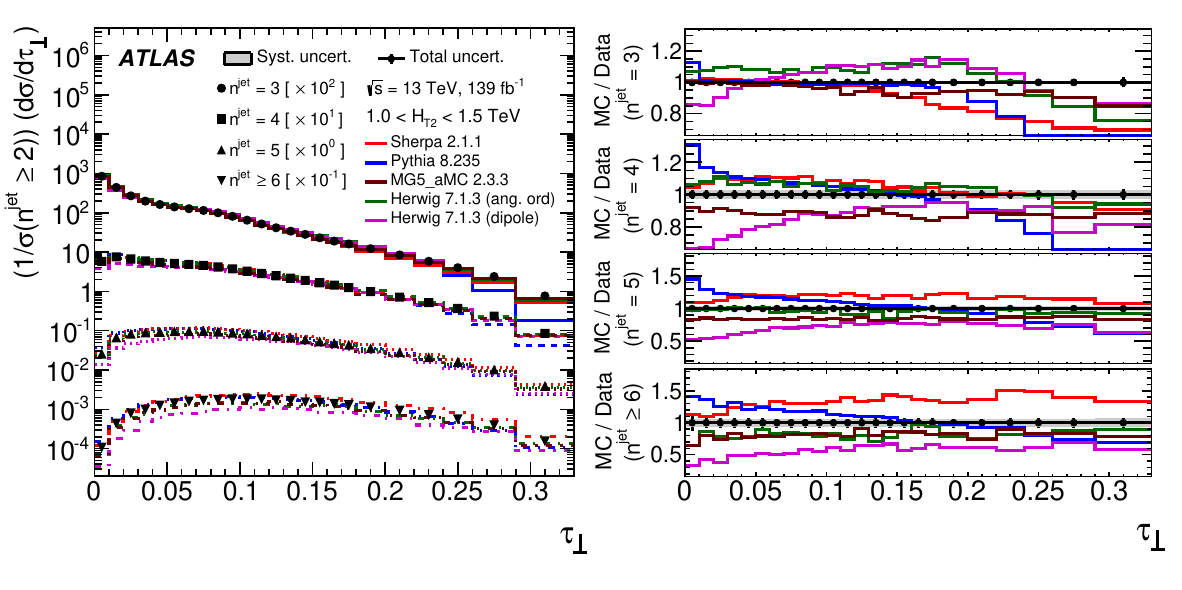}\\
\vspace{-0.5cm}
\includegraphics[width=0.98\textwidth,height=0.465\textwidth]{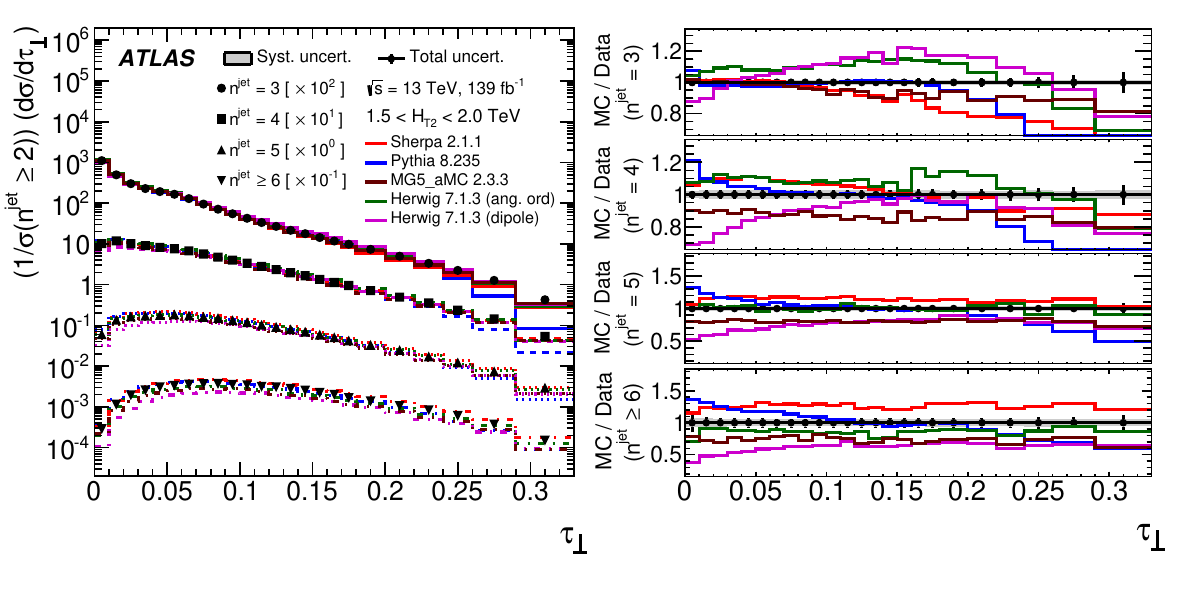}\\
\vspace{-0.5cm}
\includegraphics[width=0.98\textwidth,height=0.465\textwidth]{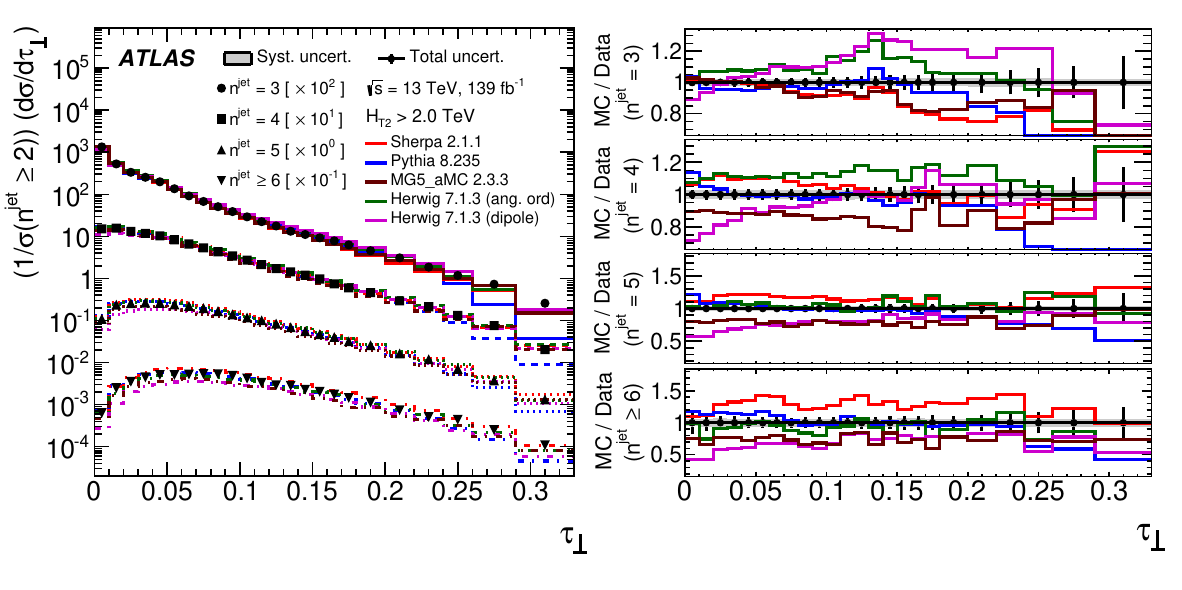}\\
\vspace{-0.55cm}
\caption{Comparison between data and MC simulation as a function of the transverse thrust $\tau_{\perp}$~ (see Eq.~\ref{eq:tauT}) for different jet multiplicities and energy scales. For illustration purposes, the corresponding differential cross section for each jet multiplicity is multiplied by 10$^{2}$ ($n^{\textrm{jet}}$ = 3), 10$^{1}$ ($n^{\textrm{jet}}$ = 4), 10$^{0}$ ($n^{\textrm{jet}}$ = 5), 10$^{-1}$ ($n^{\textrm{jet}}\geq$ 6). The right panels show the ratios between the MC and the data distributions. The error bars show the total uncertainty (statistical and systematic added in quadrature) and the grey bands in the right panels show the systematic uncertainty.}
\label{fig:xsec_transThrust}
\end{figure}
 
\begin{figure}[H]
\centering
\vspace{-0.4cm}
\includegraphics[width=0.98\textwidth,height=0.465\textwidth]{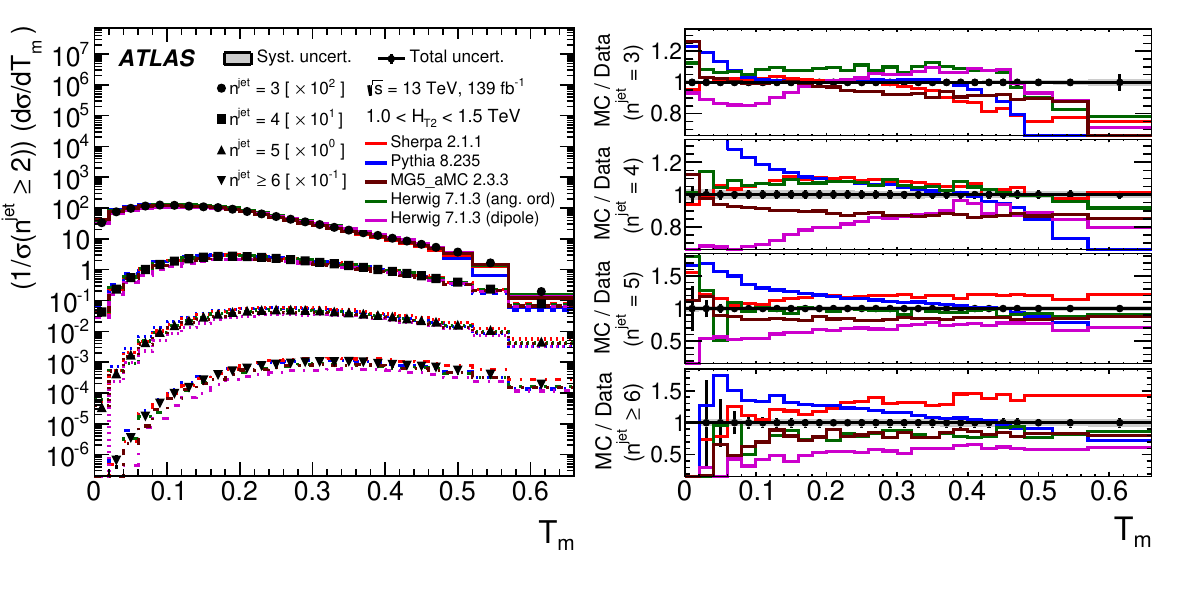}\\
\vspace{-0.5cm}
\includegraphics[width=0.98\textwidth,height=0.465\textwidth]{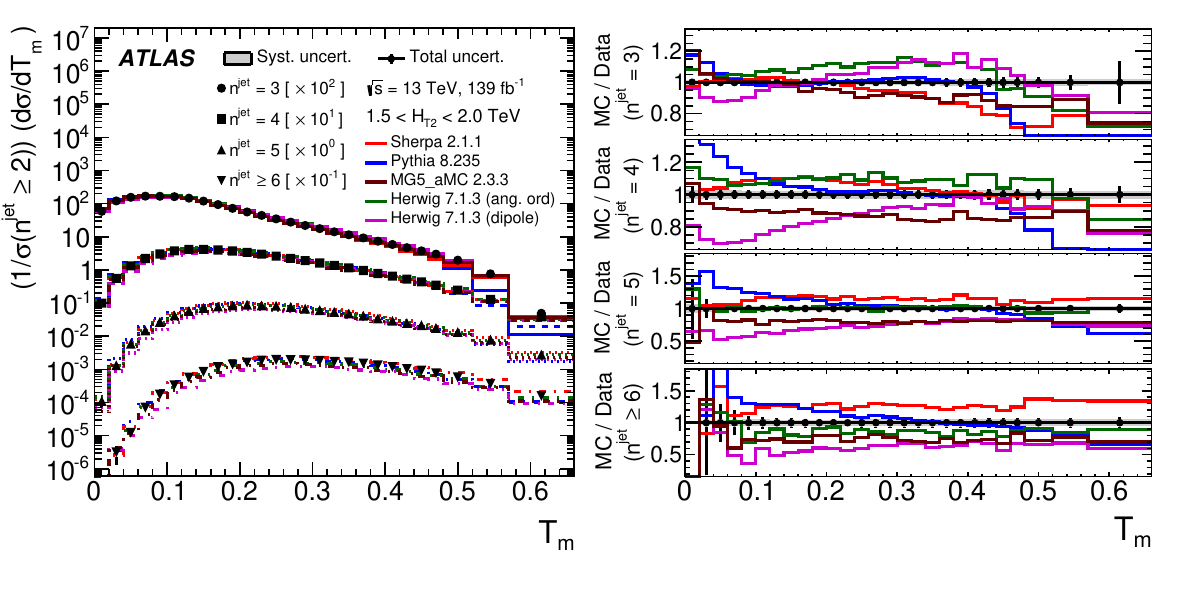}\\
\vspace{-0.5cm}
\includegraphics[width=0.98\textwidth,height=0.465\textwidth]{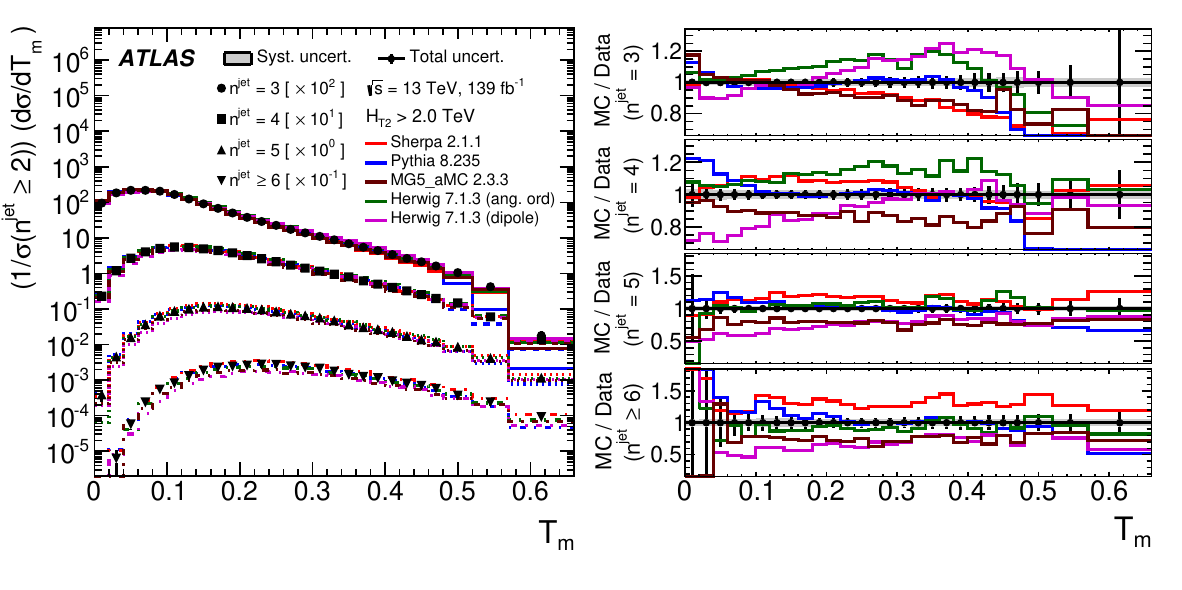}\\
\vspace{-0.55cm}
\caption{Comparison between data and MC predictions as a function of the transverse minor $T_\text{m}$~(see Eq.~\ref{eq:tauT}) for different jet multiplicities and energy scales. For illustration purposes, the corresponding differential cross section for each jet multiplicity is multiplied by 10$^{2}$ ($n^{\textrm{jet}}$ = 3), 10$^{1}$ ($n^{\textrm{jet}}$ = 4), 10$^{0}$ ($n^{\textrm{jet}}$ = 5), 10$^{-1}$ ($n^{\textrm{jet}}\geq$ 6). The right panels show the ratios between the MC and the data distributions. The error bars show the total uncertainty (statistical and systematic added in quadrature) and the grey bands in the right panels show the systematic uncertainty.}
\label{fig:xsec_transMinor}
\end{figure}
 
\begin{figure}[H]
\centering
\vspace{-0.4cm}
\includegraphics[width=0.98\textwidth,height=0.465\textwidth]{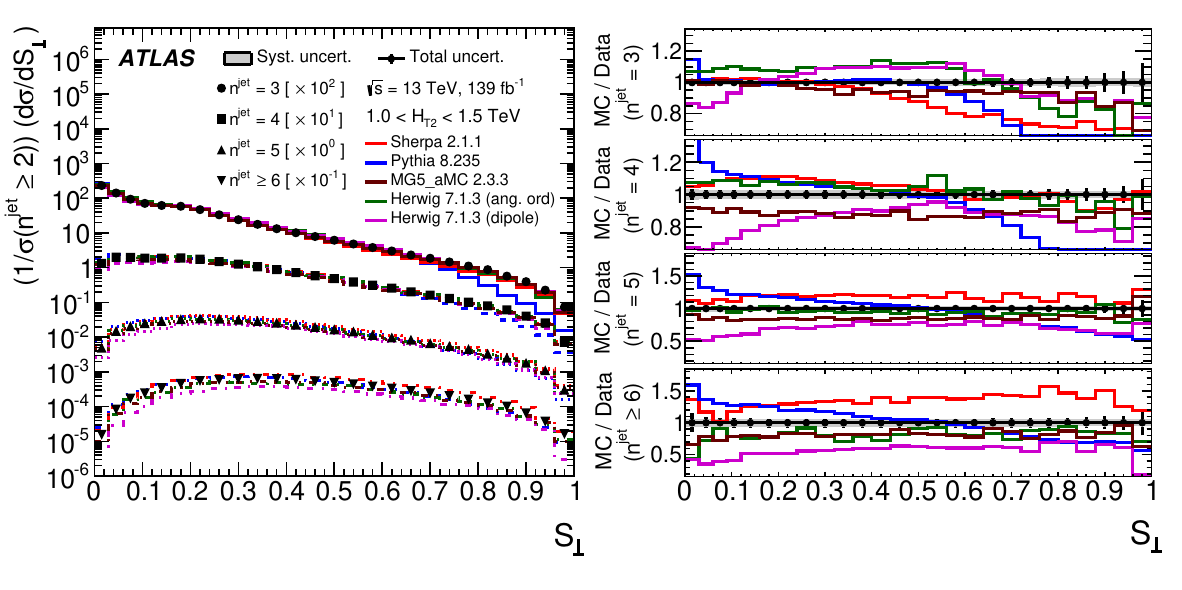}\\
\vspace{-0.5cm}
\includegraphics[width=0.98\textwidth,height=0.465\textwidth]{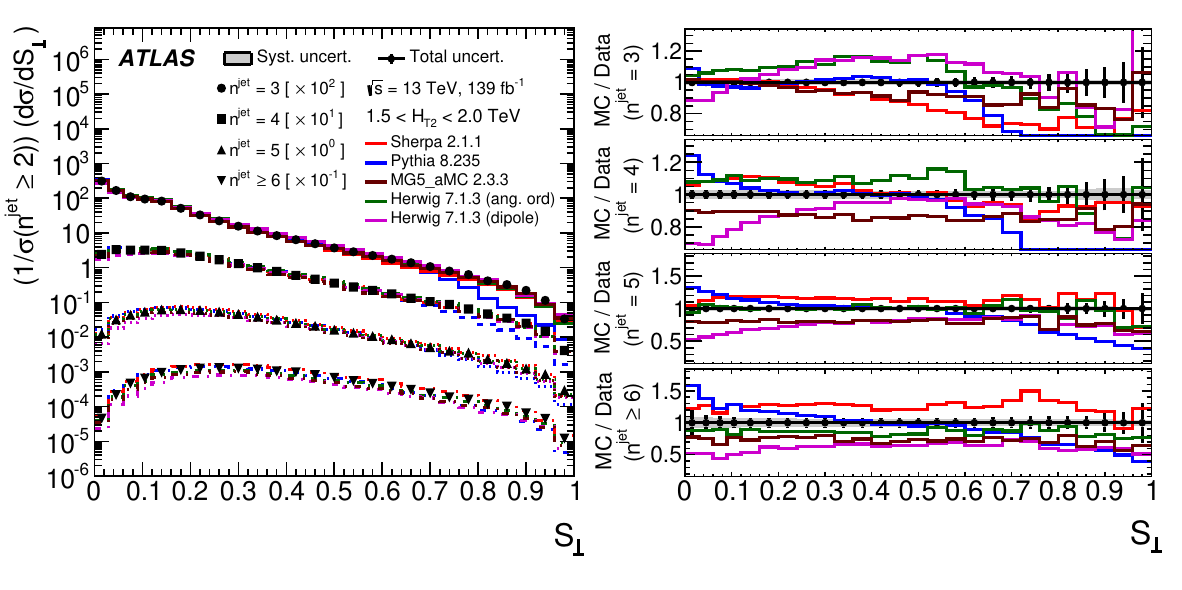}\\
\vspace{-0.5cm}
\includegraphics[width=0.98\textwidth,height=0.465\textwidth]{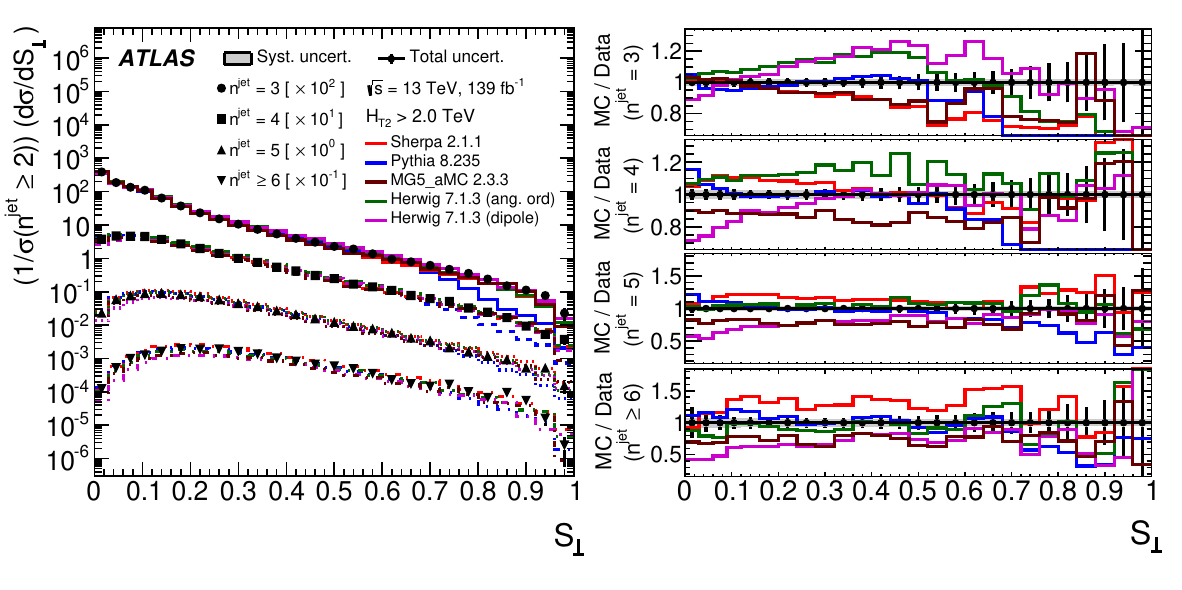}\\
\vspace{-0.55cm}
\caption{Comparison between data and MC predictions as a function of the transverse sphericity S$_{\perp}$~(see Eq.~\ref{eq:ST}) for different jet multiplicities and energy scales. For illustration purposes, the corresponding differential cross section for each jet multiplicity is multiplied by 10$^{2}$ ($n^{\textrm{jet}}$ = 3), 10$^{1}$ ($n^{\textrm{jet}}$ = 4), 10$^{0}$ ($n^{\textrm{jet}}$ = 5), 10$^{-1}$ ($n^{\textrm{jet}}\geq$ 6). The right panels show the ratios between the MC and the data distributions. The error bars show the total uncertainty (statistical and systematic added in quadrature) and the grey bands in the right panels show the systematic uncertainty.}
\label{fig:xsec_transSphericity}
\end{figure}
 
\begin{figure}[H]
\centering
\vspace{-0.4cm}
\includegraphics[width=0.98\textwidth,height=0.465\textwidth]{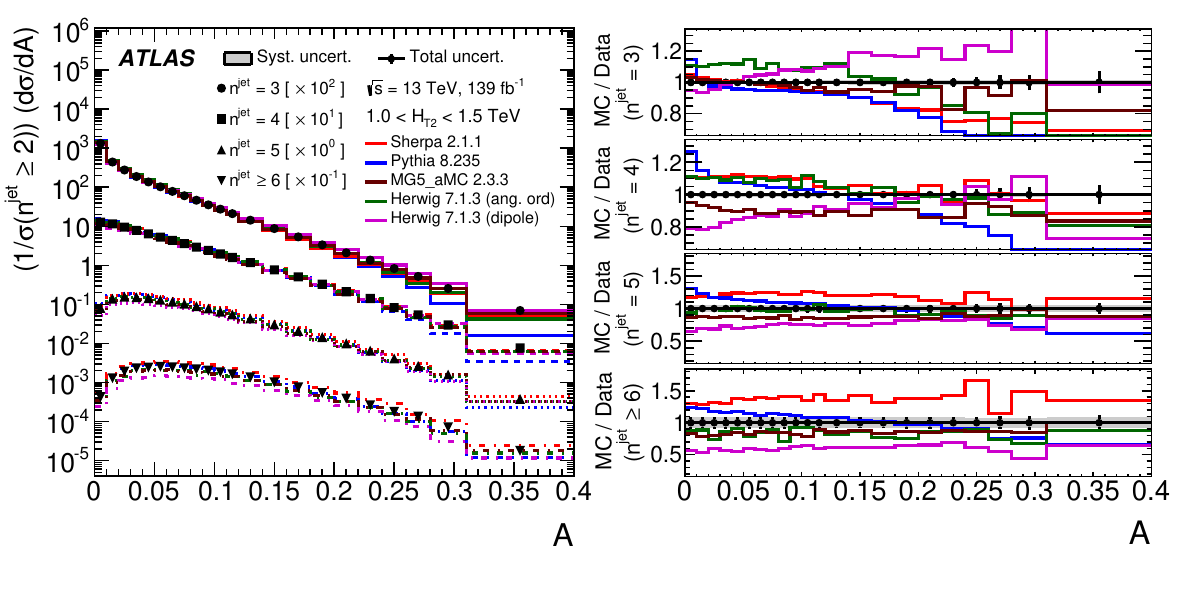}\\
\vspace{-0.5cm}
\includegraphics[width=0.98\textwidth,height=0.465\textwidth]{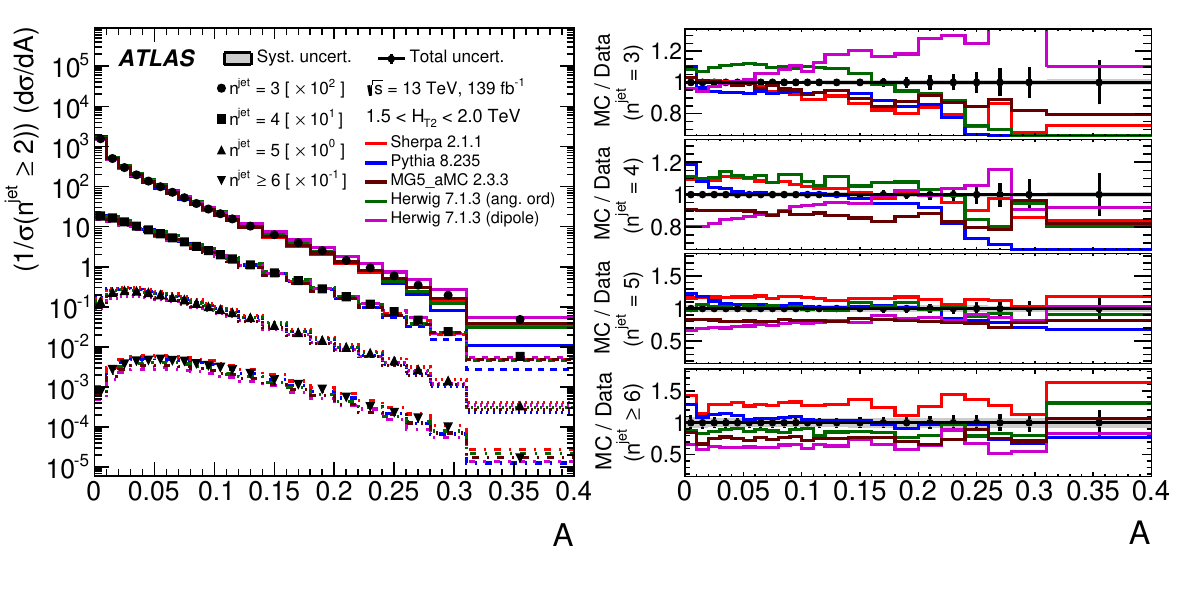}\\
\vspace{-0.5cm}
\includegraphics[width=0.98\textwidth,height=0.465\textwidth]{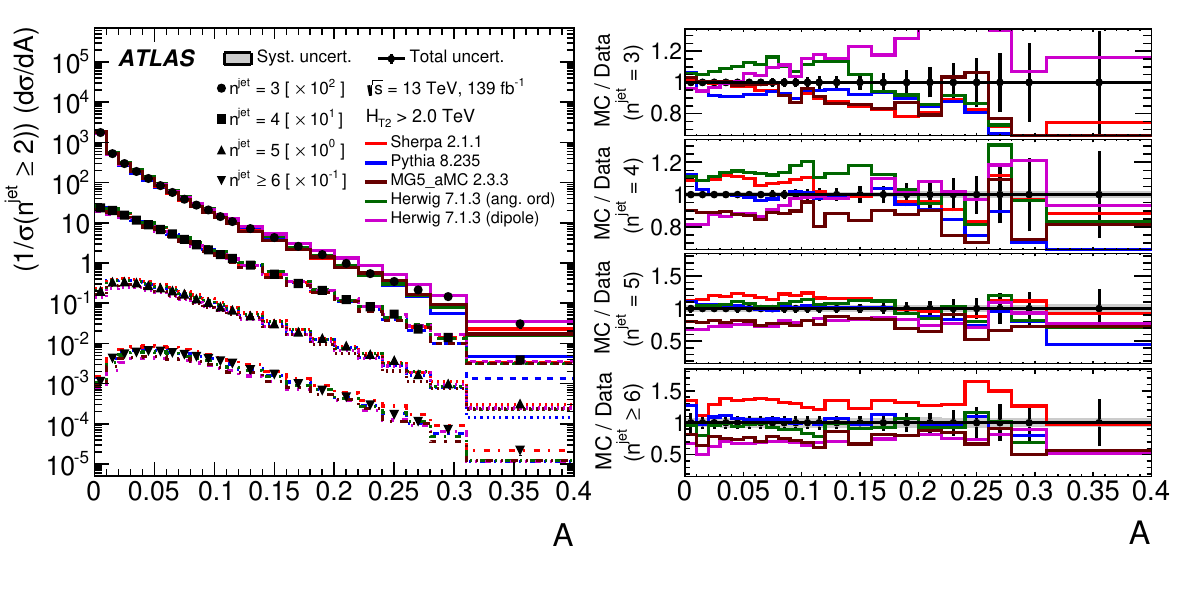}\\
\vspace{-0.55cm}
\caption{Comparison between data and MC predictions as a function of the aplanarity $A$~(see Eq.~\ref{eq:linSpher}) for different jet multiplicities and energy scales. For illustration purposes, the corresponding differential cross section for each jet multiplicity is multiplied by 10$^{2}$ ($n^{\textrm{jet}}$ = 3), 10$^{1}$ ($n^{\textrm{jet}}$ = 4), 10$^{0}$ ($n^{\textrm{jet}}$ = 5), 10$^{-1}$ ($n^{\textrm{jet}}\geq$ 6). The right panels show the ratios between the MC and the data distributions. The error bars show the total uncertainty (statistical and systematic added in quadrature) and the grey bands in the right panels show the systematic uncertainty.}
\label{fig:xsec_aplanarity}
\end{figure}
 
\begin{figure}[H]
\centering
\vspace{-0.4cm}
\includegraphics[width=0.98\textwidth,height=0.465\textwidth]{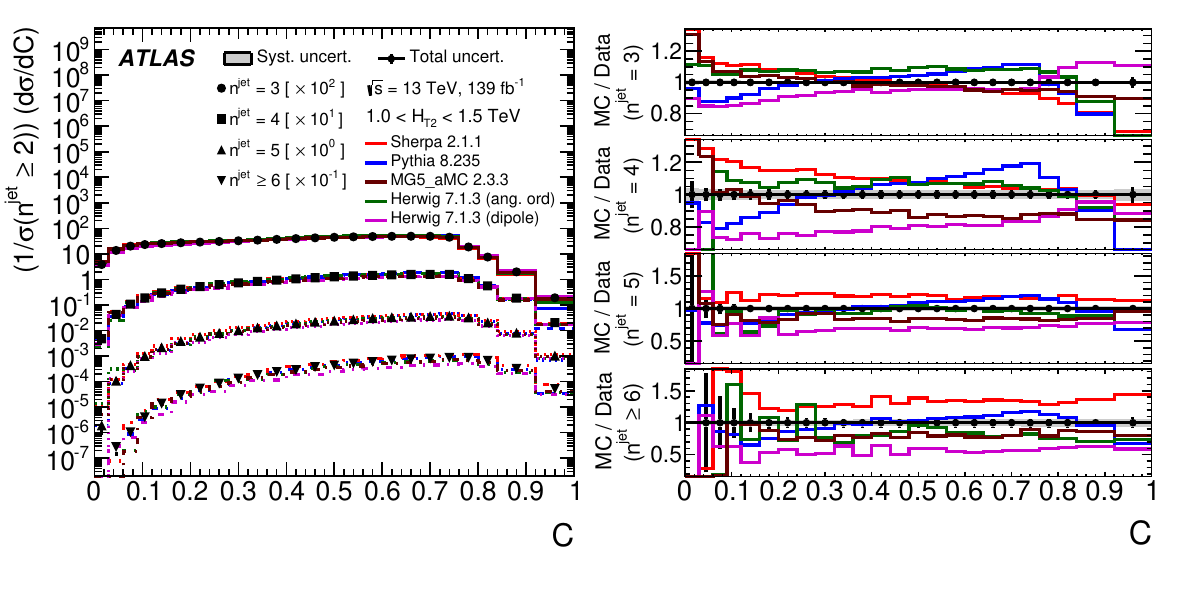}\\
\vspace{-0.5cm}
\includegraphics[width=0.98\textwidth,height=0.465\textwidth]{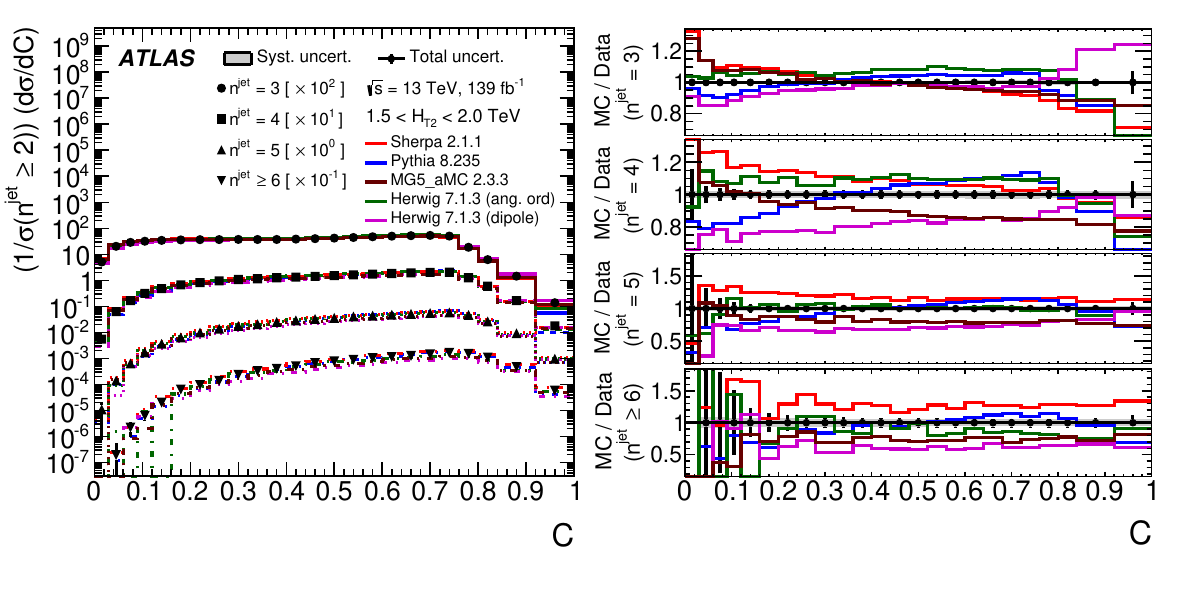}\\
\vspace{-0.5cm}
\includegraphics[width=0.98\textwidth,height=0.465\textwidth]{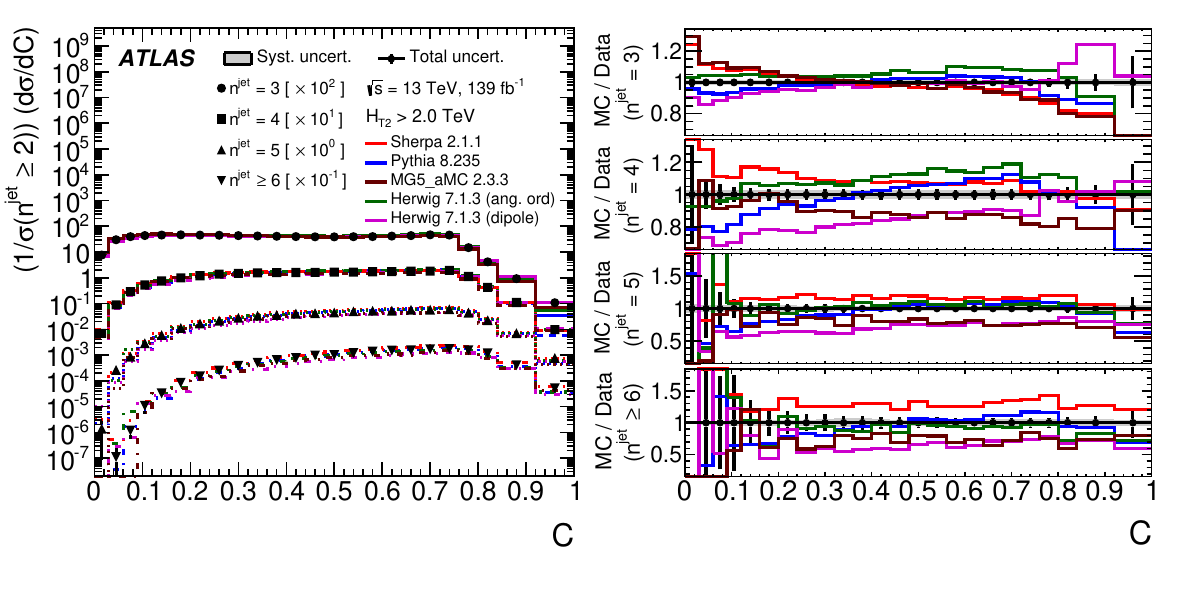}\\
\vspace{-0.55cm}
\caption{Comparison between data and MC predictions as a function of $C$~(see Eq.~\ref{eq:C}) for different jet multiplicities and energy scales. For illustration purposes, the corresponding differential cross section for each jet multiplicity is multiplied by 10$^{2}$ ($n^{\textrm{jet}}$ = 3), 10$^{1}$ ($n^{\textrm{jet}}$ = 4), 10$^{0}$ ($n^{\textrm{jet}}$ = 5), 10$^{-1}$ ($n^{\textrm{jet}}\geq$ 6). The right panels show the ratios between the MC and the data distributions. The error bars show the total uncertainty (statistical and systematic added in quadrature) and the grey bands in the right panels show the systematic uncertainty.}
\label{fig:xsec_C}
\end{figure}
 
\begin{figure}[H]
\centering
\vspace{-0.4cm}
\includegraphics[width=0.98\textwidth,height=0.465\textwidth]{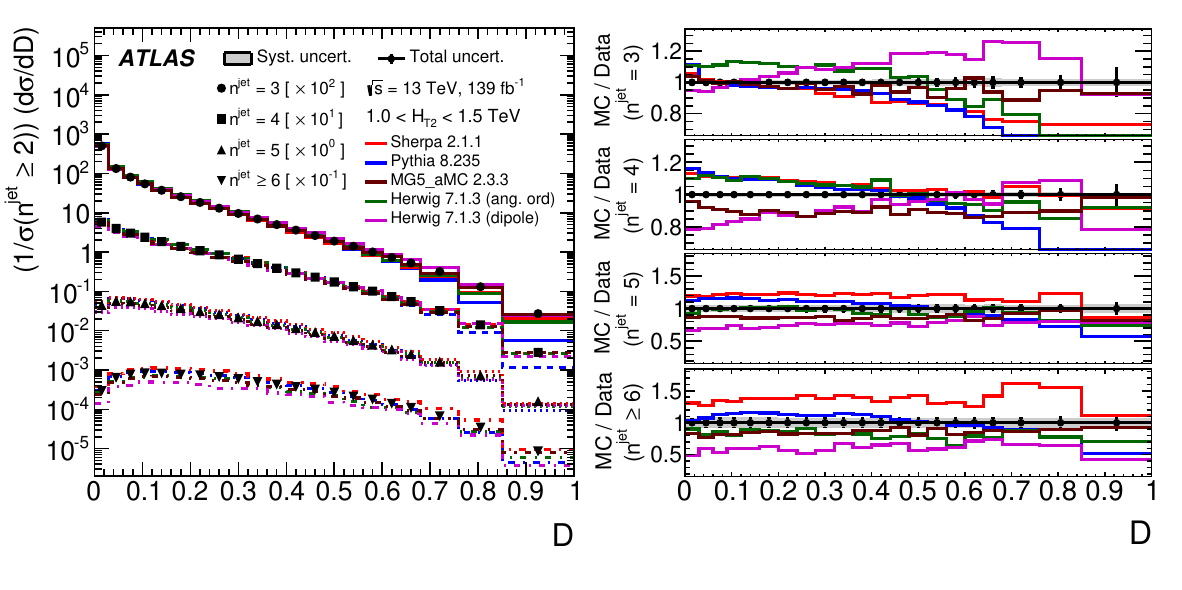}\\
\vspace{-0.5cm}
\includegraphics[width=0.98\textwidth,height=0.465\textwidth]{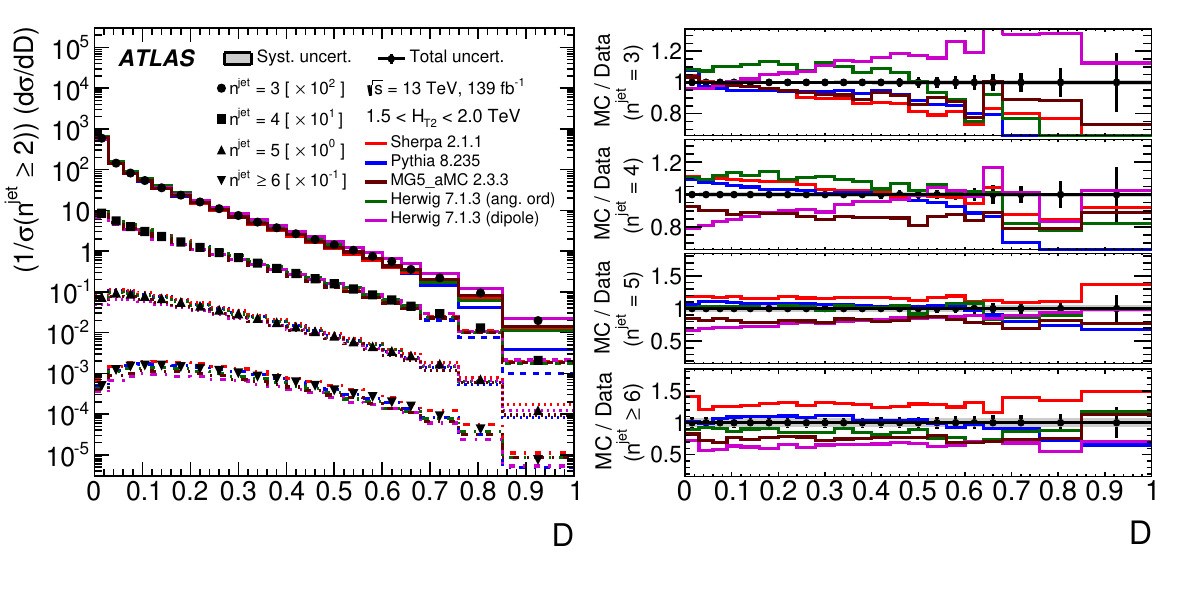}\\
\vspace{-0.5cm}
\includegraphics[width=0.98\textwidth,height=0.465\textwidth]{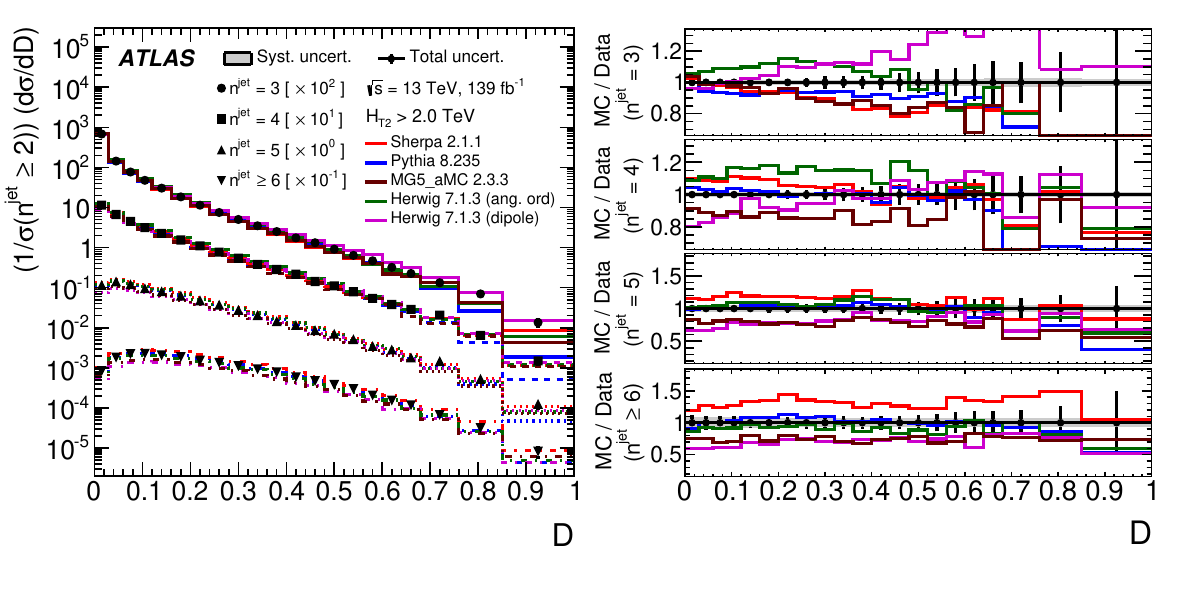}\\
\vspace{-0.55cm}
\caption{Comparison between data and MC predictions as a function of $D$~(see Eq.~\ref{eq:D}) for different jet multiplicities and energy scales. For illustration purposes, the corresponding differential cross section for each jet multiplicity is multiplied by 10$^{2}$ ($n^{\textrm{jet}}$ = 3), 10$^{1}$ ($n^{\textrm{jet}}$ = 4), 10$^{0}$ ($n^{\textrm{jet}}$ = 5), 10$^{-1}$ ($n^{\textrm{jet}}\geq$ 6). The right panels show the ratios between the MC and the data distributions. The error bars show the total uncertainty (statistical and systematic added in quadrature) and the grey bands in the right panels show the systematic uncertainty.}
\label{fig:xsec_D}
\end{figure}
\clearpage
\clearpage
\section{Summary and conclusions}
\label{sec:conclusions}
A measurement of event shapes in multijet final states is presented. The data correspond to 139~$\ifb$ of $pp$ collisions at $\sqrt{s} = 13$ TeV, recorded by the ATLAS detector at the LHC during the period 2015--2018.\\
\newline
The measurements are classified in bins of the jet multiplicity and energy scale, and six event-shape observables are measured and corrected to particle level through an unfolding procedure. The distributions of event-shape observables are normalised to the inclusive two-jet cross section 
to reduce correlated experimental uncertainties. The main uncertainties in the measurement are due to the jet calibration procedure and the MC model used in the unfolding, and the total uncertainty is at the percent level.\\
\newline
The measurements are compared with MC predictions with matrix elements calculated up to different orders in perturbative QCD, matched to different parton shower algorithms. All the predictions qualitatively describe the main features of the data, but none of them gives a good description of all distributions within the experimental uncertainties. The discrepancies between data and all the MC samples investigated show that further refinement of the current MC predictions is needed to describe the data in some regions, particularly at high jet multiplicities. Moreover, these discrepancies show that these data provide a powerful testing ground for the understanding of the strong interaction at high energies.
\clearpage
\clearpage

We thank CERN for the very successful operation of the LHC, as well as the
support staff from our institutions without whom ATLAS could not be
operated efficiently.
 
We acknowledge the support of
ANPCyT, Argentina;
YerPhI, Armenia;
ARC, Australia;
BMWFW and FWF, Austria;
ANAS, Azerbaijan;
SSTC, Belarus;
CNPq and FAPESP, Brazil;
NSERC, NRC and CFI, Canada;
CERN;
ANID, Chile;
CAS, MOST and NSFC, China;
Minciencias, Colombia;
MSMT CR, MPO CR and VSC CR, Czech Republic;
DNRF and DNSRC, Denmark;
IN2P3-CNRS and CEA-DRF/IRFU, France;
SRNSFG, Georgia;
BMBF, HGF and MPG, Germany;
GSRI, Greece;
RGC and Hong Kong SAR, China;
ISF and Benoziyo Center, Israel;
INFN, Italy;
MEXT and JSPS, Japan;
CNRST, Morocco;
NWO, Netherlands;
RCN, Norway;
MEiN, Poland;
FCT, Portugal;
MNE/IFA, Romania;
JINR;
MES of Russia and NRC KI, Russian Federation;
MESTD, Serbia;
MSSR, Slovakia;
ARRS and MIZ\v{S}, Slovenia;
DSI/NRF, South Africa;
MICINN, Spain;
SRC and Wallenberg Foundation, Sweden;
SERI, SNSF and Cantons of Bern and Geneva, Switzerland;
MOST, Taiwan;
TAEK, Turkey;
STFC, United Kingdom;
DOE and NSF, United States of America.
In addition, individual groups and members have received support from
BCKDF, CANARIE, Compute Canada and CRC, Canada;
COST, ERC, ERDF, Horizon 2020 and Marie Sk{\l}odowska-Curie Actions, European Union;
Investissements d'Avenir Labex, Investissements d'Avenir Idex and ANR, France;
DFG and AvH Foundation, Germany;
Herakleitos, Thales and Aristeia programmes co-financed by EU-ESF and the Greek NSRF, Greece;
BSF-NSF and GIF, Israel;
Norwegian Financial Mechanism 2014-2021, Norway;
NCN and NAWA, Poland;
La Caixa Banking Foundation, CERCA Programme Generalitat de Catalunya and PROMETEO and GenT Programmes Generalitat Valenciana, Spain;
G\"{o}ran Gustafssons Stiftelse, Sweden;
The Royal Society and Leverhulme Trust, United Kingdom.
 
The crucial computing support from all WLCG partners is acknowledged gratefully, in particular from CERN, the ATLAS Tier-1 facilities at TRIUMF (Canada), NDGF (Denmark, Norway, Sweden), CC-IN2P3 (France), KIT/GridKA (Germany), INFN-CNAF (Italy), NL-T1 (Netherlands), PIC (Spain), ASGC (Taiwan), RAL (UK) and BNL (USA), the Tier-2 facilities worldwide and large non-WLCG resource providers. Major contributors of computing resources are listed in Ref.~\cite{ATL-SOFT-PUB-2021-003}.
 
\clearpage

\newpage
\printbibliography
 
\clearpage
 
\begin{flushleft}
\hypersetup{urlcolor=black}
{\Large The ATLAS Collaboration}

\bigskip

\AtlasOrcid[0000-0002-6665-4934]{G.~Aad}$^\textrm{\scriptsize 102}$,    
\AtlasOrcid[0000-0002-5888-2734]{B.~Abbott}$^\textrm{\scriptsize 128}$,    
\AtlasOrcid[0000-0002-7248-3203]{D.C.~Abbott}$^\textrm{\scriptsize 103}$,    
\AtlasOrcid[0000-0002-2788-3822]{A.~Abed~Abud}$^\textrm{\scriptsize 36}$,    
\AtlasOrcid[0000-0002-1002-1652]{K.~Abeling}$^\textrm{\scriptsize 53}$,    
\AtlasOrcid[0000-0002-2987-4006]{D.K.~Abhayasinghe}$^\textrm{\scriptsize 94}$,    
\AtlasOrcid[0000-0002-8496-9294]{S.H.~Abidi}$^\textrm{\scriptsize 167}$,    
\AtlasOrcid[0000-0002-8279-9324]{O.S.~AbouZeid}$^\textrm{\scriptsize 40}$,    
\AtlasOrcid{N.L.~Abraham}$^\textrm{\scriptsize 156}$,    
\AtlasOrcid[0000-0001-5329-6640]{H.~Abramowicz}$^\textrm{\scriptsize 161}$
% ...
\bigskip
\\

% ...
$^{36}$CERN, Geneva; Switzerland.\\
% ...
$^{40}$Niels Bohr Institute, University of Copenhagen, Copenhagen; Denmark.\\
% ...
$^{53}$II. Physikalisches Institut, Georg-August-Universit\"{a}t G\"ottingen, G\"ottingen; Germany.\\
% ...
$^{94}$Department of Physics, Royal Holloway University of London, Egham; United Kingdom.\\
% ...
$^{102}$CPPM, Aix-Marseille Universit\'e, CNRS/IN2P3, Marseille; France.\\
$^{103}$Department of Physics, University of Massachusetts, Amherst MA; United States of America.\\
% ...
$^{128}$Homer L. Dodge Department of Physics and Astronomy, University of Oklahoma, Norman OK; United States of America.\\
% ...
$^{156}$Department of Physics and Astronomy, University of Sussex, Brighton; United Kingdom.\\
% ...
$^{161}$Raymond and Beverly Sackler School of Physics and Astronomy, Tel Aviv University, Tel Aviv; Israel.\\
% ...
$^{167}$Department of Physics, University of Toronto, Toronto ON; Canada.\\
% ...

\end{flushleft}


\end{document}